\documentclass[epj,numbook]{svjour}
\usepackage{epsfig,rotate}
\newcommand{\vecj}{\mbox{\bf \j}}
\newcommand{\scj}{\mbox{\j}}
\begin{document}
\title{Non-equilibrium behavior at a liquid-gas critical point} 


\author{Jaime E. Santos\inst{1} and Uwe C. T\"auber\inst{2}}

%
%

\institute{
Hahn-Meitner Institut, Abt. SF5, Glienicker Str. 100, 14109 Berlin, 
Germany\\ \email{santos@hmi.de}
\and
Physics Department, Virginia Polytechnic Institute and
State University, Blacksburg, VA 24061-0435, USA\\ 
\email{tauber@vt.edu} } 
\date{}
%
\abstract{\rm
Second-order phase transitions in a non-equilibrium liquid-gas 
model with reversible mode couplings, i.e., model H for binary-fluid 
critical dynamics, are studied using dynamic field theory and the 
renormalization group. The system is driven out of equilibrium 
either by considering different values for the noise strengths in 
the Langevin equations describing the evolution of the dynamic 
variables (effectively placing these at different temperatures), 
or more generally by allowing for anisotropic noise strengths, i.e., 
by constraining the dynamics to be at different temperatures in 
$d_\parallel$- and $d_\perp$-dimensional subspaces, respectively. 
In the first, {\em isotropic} case, we find one infrared-stable and 
one unstable renormalization group fixed point. At the stable fixed 
point, detailed balance is dynamically restored, with the two noise 
strengths becoming asymptotically equal. The ensuing critical 
behavior is that of the standard equilibrium model H. At the novel 
unstable fixed point, the temperature ratio for the dynamic variables 
is renormalized to infinity, resulting in an effective decoupling 
between the two modes. We compute the critical exponents at this new
fixed point to one-loop order. For model H with spatially 
{\em anisotropic} noise, we observe a critical softening only in the 
$d_\perp$-dimensional sector in wave vector space with lower noise 
temperature. The ensuing effective two-temperature model H does not 
have any stable fixed point in any physical dimension, at least to 
one-loop order. We obtain formal expressions for the novel critical 
exponents in a double expansion about the upper critical dimension 
$d_c = 4 - d_\parallel$ and with respect to $d_\parallel$, i.e., 
about the equilibrium theory.
} 
\maketitle

\section{Introduction}
 \label{introd}

The theory of equilibrium dynamics for critical phenomena has identified 
a set of universality classes which describe the long-time, 
long-wavelength behavior of systems in the vicinity of critical points 
\cite{hohhal}. The situation is in this respect similar to that found 
in the study of static critical phenomena. However, whereas in the
static case the nature of the universality classes is determined solely 
by the nature of the interactions that fixes the dimensionality $n$ of 
the order parameter space as well as the effective spatial dimension $d$
for its fluctuations, in the case of dynamical critical phenomena the 
corresponding universality classes are not simply characterized by the 
static or dynamic interactions which exist between the relevant system
variables, but they also crucially depend on the conservation laws 
which are implemented by the dynamics and on the very existence of the
detailed balance constraint that guarentees relaxation to thermal
equilibrium in the long-time limit. 
 
At the quantitative level, the equilibrium dynamics of critical systems 
is usually formulated in terms of coupled non-linear Langevin equations. 
Such equations typically include a dissipative term which involves the 
Hamiltonian describing the static critical properties and a Gaussian 
white noise term, which mimics thermal fluctuations and the random 
forces originating from couplings to fast, non-critical non-conserved
modes. Furthermore, the equations of motion may also include purely 
reversible mode-coupling terms which represent dynamical interactions 
between the relevant slow system variables. In this framework, the 
conditions for the existence of detailed balance are (i) the Einstein 
relation between the relaxation constants and the noise strengths, and 
(ii) the condition that the probability current associated with 
reversible mode-coupling terms be divergence-free in the space of slow 
dynamic variables \cite{intcon}. The existence of conservation laws in 
turn fixes the precise form of the relaxation terms \cite{nonoether}.
If the coupled system of Langevin equations obeys these two conditions, 
it can be shown, e.g., by means of the associated Fokker-Planck 
equation, that the long-time steady state of the dynamics is indeed
characterized by a Gibbsian probability distribution, with precisely
the Hamiltonian that describes the static critical properties of the 
system. These conditions also insure that the dynamic susceptibilities 
reduce to the static ones in the limit of zero frequency, and imply the
validity of fluctuation-dissipation relations between dynamical
correlation functions and the dissipative parts of the response. 

On a more general footing, one can also consider the study of dynamical 
phenomena which are truly non-equi\-li\-brium in the sense that they do 
not possess a steady state described by a Gibbs distribution, of which 
the best-known example is perhaps the Kardar-Parisi-Zhang equation (for
$d > 1$) \cite{kapezt}, which describes the curvature-driven epitaxial 
growth of a surface. Other prominent examples are driven diffusive 
systems \cite{bearev}, models of driven interfaces and growing surfaces 
\cite{intrev}, depinning transitions \cite{deprev}, and phase 
transitions from active to absorbing states \cite{dirper}, e.g., in 
diffusion-limited chemical reactions.

A question which lies at the interface between these two subjects, i.e., 
between the study of systems with equilibrium and far-from-equilibrium 
dynamics, concerns the asymptotic scaling properties of a critical 
system originally in thermal equilibrium, which is however subjected to 
a perturbation that violates the detailed balance conditions. This issue
is also relevant in an experimental context, because the maintenance of 
thermodynamic equilibrium in a critical system during an experiment is a 
non-trivial task, as due to critical slowing-down the relaxation times 
become very long. For the interpretation of the experimental results, it 
might thus be important to know whether the dynamical system will 
eventually be driven to a genuine non-equilibrium fixed point which is 
characterized by scaling exponents distinct from the original 
equilibrium critical behavior. A priori, a perturbation from thermal 
equilibrium may either imply a violation of condition (i) or of 
condition (ii) above. This latter case includes, for example, driven 
lattice gases in which the terms added to the Langevin equation 
modelling the equilibrium diffusive dynamics stem from a global gradient 
such as an energy or mass current through the system \cite{bearev}. The 
violation of condition (i) corresponds, on the other hand, to the 
coupling of the system to a {\em local} energy gradient. This type of 
perturbation was the subject of study in 
Refs.~\cite{bearoy,uwezol,uwejai,uwevam}, where the consequences of the 
violation of the Einstein relations in several equilibrium dynamical 
models, largely describing magnetic systems, were investigated within 
the framework of the dynamical renormalization group. Specifically, two 
generic types of detailed-balance violations were studied there, namely 
(a) coupling the order parameter and additional conserved quantities to 
heat baths with different temperatures, and (b) allowing for spatially 
anisotropic noise correlations for conserved variables.

In this paper, we extend the above studies 
\cite{bearoy,uwezol,uwejai,uwevam} to a model which describes the 
dynamics of a liquid-gas phase transition, or the phase separation of a 
binary liquid, at its critical point. This is the so-called model H in 
the terminology of Ref.~\cite{hohhal}, which incorporates the 
interaction of the conserved scalar order parameter, a linear 
combination of the mass and energy densities, namely essentially the 
free energy density, with the conserved transverse mass
current vector via a reversible mode-coupling term \cite{modelH}. 
These two fields are sufficient to describe the critical dynamics 
although critical effects can also be seen in other modes, such as the 
sound mode \cite{folk1}. Furthermore, a renormalization group analysis
was applied to investigate non-universal properties and crossover 
behavior in the equilibrium dynamics \cite{folk2}. The original 
approaches to the equilibrium critical dynamics of model H utilized 
self-consistent mode coupling theory \cite{kawas}. More recently, 
Patashinski performed a linear analysis of the equations of motion to 
study the effect of perturbations which induce a non-uniform 
non-equilibrium stationary state \cite{patash}.

We use the response functional formalism \cite{dynfun} to express the 
dynamical equations as a path integral which represents a very 
convenient form to develop the perturbation expansion and subsequent 
renormalization group analysis. We compute the beta functions of the 
theory to one-loop order in the isotropic case where the two dynamical 
variables are coupled to heat baths at different temperatures and also 
in the case where we allow for spatially anisotropic noise correlations 
for these conserved fields. In the first situation, we obtain one stable 
and one unstable fixed point. The stable fixed point is just the  
ordinary equilibrium fixed point, for which detailed balance is 
dynamically restored, and in whose vicinity the two noise temperatures
become equal. On the other hand, at the genuinely non-equilibrium, but
unstable fixed point, the ratio between the temperature of the noise 
coupled to the order parameter and that of the noise coupled to the 
conserved current is renormalized to infinity. In this limiting case,
the critical exponents of the conserved order parameter are simply 
those of the decoupled diffusive model B. This is due to the existence 
of a unidirectional random heat flow from the order parameter heat bath 
to the mass current thermal reservoir which renders the effect of the 
mode-coupling terms in the dynamics of the order parameter negligible. 
This situation is quite analogous to that found in the earlier study of 
the non-equilibrium S\'asvari--Schwabl--Sz\'epfalusy (SSS) model 
\cite{uwezol}, although there the order parameter in not conserved, 
and hence follows model A dynamics in the corresponding limiting case. 
Moreover, owing to the fact that a conserved critical field always 
relaxes much slower than a diffusive mode, here we find no fixed point 
that would describe the above temperature ratio scaling to zero. 

When allowing for spatially anisotropic noise, we observe a softening 
of the dynamics only in the $d_\perp$-dimensio\-nal sector in wave 
vector space with lower noise temperature. We show that the renormalized 
coupling constants diverge as the dimension of the subspace with higher 
noise temperature approaches $d_\parallel=0.838454$; thus, at least to
one-loop order, there exists no finite fixed point for any physical 
value of $d_\parallel$. We then obtain formal expressions for the novel 
critical exponents in a double expansion about the upper critical 
dimension $d_c = 4 - d_\parallel$ and with respect to $d_\parallel$, 
i.e., about the equilibrium theory. Here again the results are similar 
to those we have previously obtained in the study of the two-temperature
non-equilibrium model J \cite{uwejai}. These conclusions are thus in 
line with our earlier studies and with previous observations that 
whereas models with a non-conserved order parameter are quite robust 
against violations of the detailed balance conditions, models with 
conserved dynamics seem to be extremely sensitive to this type of 
perturbations provided they are rendered {\em anisotropic} \cite{uwevam}.  

The structure of this paper is as follows: In section \ref{modeqs} we 
present the coupled non-linear Langevin equations which define our two 
non-equilibrium versions of model H, namely the isotropic case where 
the two dynamic variables are at different temperatures, and the 
situation where we allow for anisotropic noise correlations, and we 
show how these models can be mapped to a dynamic field theory. The 
action of these functionals is then separated into a Gaussian part and 
a non-linear contribution which is to be treated in a perturbation 
expansion. In section \ref{isoH}, we present the renormalization group 
analysis of the isotropic non-equilibrium model H to one-loop order and 
discuss its behavior near the renormalization group fixed points. In 
section \ref{anisoH}, after introducing spatially anisotropic noise 
correlations, we derive the renormalization group flow equations for the 
resulting effective two-temperature model H, and we compute the values 
of the critical exponents in a formal double expansion about the 
equilibrium theory, also to one-loop order. We demonstrate that there 
exists no one-loop renormalizationg group fixed point in any physical 
dimension, however. Finally, in section \ref{concl} we present our 
conclusions.

\section{Model equations}
 \label{modeqs}

In this section, we briefly outline the basic model equations for 
our isotropic and anisotropic non-equilibrium generalizations of 
model H. Following the equilibrium theory \cite{modelH}, we consider 
a second-order phase transition for the scalar order parameter 
$\psi_0(\vec{x},t)=e(\vec{x},t)+(\bar\mu-T\bar{s})$ $\rho(\vec{x},t)$ 
at a liquid-gas (or binary-fluid) critical point, where $e(\vec{x},t)$ 
is the energy density, $\rho(\vec{x},t)$ the mass density and 
$\bar{\mu}$, $T$ and $\bar{s}$ are, respectively, the chemical potential, 
the temperature and the entropy at equilibrium (we denote unrenormalized 
quantities by the subscript `$0$'). This order parameter is dynamically 
coupled to the transverse mass current $\vecj_0(\vec{x},t)$, which 
satisfies $\nabla\cdot \vecj_0=0$. Notice that the longitudinal current 
$\vecj^\parallel_0$ is related to the mass density through the continuity 
equation $\partial_t\rho + \nabla\cdot \vecj^\parallel_0 = 0$. The two 
dynamic fields $\psi_0$ and $\vecj_0$ yield three hydrodynamic modes, 
namely thermal and viscous diffusion. A fourth hydrodynamic mode, i.e., 
the sound mode, is also present in a fluid, but for low momenta its 
characteristic frequency is much higher than the frequencies of the shear 
and energy modes and one may thus disregard it to first approximation. 

Next, the static critical properties of the system are described by the 
following Landau--Ginzburg--Wilson free energy functional in $d$ space 
dimensions,
\begin{eqnarray}
        H_0[\psi_0,\vecj_0] = \int \! d^dx &&\Biggl\{ {r_0 \over 2}\, 
        \psi^2_0(\vec{x})+ {1 \over 2}\,
                \left[ \nabla \psi_0(\vec{x})\right]^{2} \nonumber\\
        &&\mbox{}+ {u_0 \over 4 !}\, \psi^{4}_{0}(\vec{x})+{1\over 2}\,
        \vecj_{0}^{2}(\vec{x}) 
              \Biggr\} \ ,
 \label{hamilt}
\end{eqnarray}
where $r_0 = (T - T_c^0) / T_c^0$ denotes the relative deviation from 
the mean-field critical temperature $T_c^0$. Note that the quadratic 
term in $\vecj_{0}$ simply represents the kinetic energy of a 
stationary mass current. Since $\vecj_0$ itself is a non-critical 
variable, the coefficient in front of this term will be weakly
temperature-dependent only, and can thus be set constant in the 
vicinity of $T_c$, and eventually absorbed into $\vecj_0$.

This effective free energy (\ref{hamilt}) determines the equilibrium 
probability distribution for the vector order parameter $\psi_0$ and 
for the mass current $\vecj_0$,
\begin{equation}
        P_{\rm eq}[\psi_0,\vecj_0] = {e^{-H_0[\psi_0,\vecj_0] / k_{\rm B} T} 
                \over \int {\cal D}[\psi_0]\,{\cal D}[\vecj_0]\, 
                e^{-H_0[\psi_0,\vecj_0] / k_{\rm B} T}} \ .
 \label{eqdist}
\end{equation}
Note that it follows from (\ref{eqdist}) that at the purely static 
level the order parameter $\psi_0(\vec{x})$ and the mass current 
$\vecj_0(\vec{x})$ are completely decoupled; moreover, $\vecj_0(\vec{x})$ 
is a Gaussian variable, whose contribution to the free energy can be 
readily factored out of the functional integral. The task is then reduced 
to the computation of two independent critical exponents, e.g., $\eta$ 
and $\nu$, by means of perturbation theory with respect to the static 
non-linear coupling $u_0$ and by employing the renormalization group 
procedure, within a systematic expansion in terms of $\epsilon = 4-d$ 
about the static upper critical dimension $d_c = 4$. Here, $\eta$ 
describes the power-law decay of the order parameter correlation 
function at criticality, $\langle \psi({\vec x}) \psi({\vec x}') \rangle 
\propto 1 / |{\vec x} - {\vec x}'|^{d - 2 + \eta}$, or, equivalently, of 
the static susceptibility, $\chi({\vec q}) \propto 1 / q^{2 - \eta}$, 
and the exponent $\nu$ characterizes the divergence of the correlation 
length as $T_c$ is approached, $\xi \propto |T - T_c|^{-\nu}$. Notice 
that fluctuations also shift the true transition temperature $T_c$ 
downwards as compared to the mean-field critical temperature $T_c^0$, 
i.e., $r_{0c} = T_c - T_c^0 < 0$. Since we will need to consider
(time-dependent) correlation functions of the dynamical variables, it is 
convenient to add to the free energy functional a term involving external 
sources $h$ and $\vec A$, whereupon the (generating) functional reads
$H=H_0-\int d^dx \left[ h(\vec{x})\psi_0 (\vec{x})+\vecj_0(\vec{x})\cdot
\vec{A}(\vec{x}) \right]$.
 
With this free energy functional $H$, the coupled non-linear Langevin 
equations defining model H read \cite{modelH}
\begin{eqnarray}
        {\partial \psi_0\over \partial t} &=&
        \lambda_0\nabla^2{\delta H\over\delta \psi_0}-g_0\nabla\psi_0\cdot
         {\delta H\over\delta \vecj_0}+\eta
 \label{hoplao} \\
        &=&\lambda_0\nabla^2(r_0-\nabla^2)\psi_0+{\lambda_0 u_0\over 6}
           \nabla^2\psi^3_0 -\lambda_0\nabla^2 h \nonumber \\
        &&\mbox{}-g_0\vecj_0\cdot\nabla\psi_0
          +g_0\,\nabla\psi_0\cdot\vec{A}+\eta
        \label{hoplan}
\end{eqnarray}
and
\begin{eqnarray}
        {\partial \mbox{\bf \j}_0\over \partial t} &=&
        {\cal T}\left[
        D_0\nabla^2{\delta H\over\delta \vecj_0}+g_0\nabla\psi_0
         {\delta H\over\delta \psi_0}+\vec{\zeta}\right]
 \label{samlao} \\
        &=&  {\cal T}\left[D_0\nabla^2\vecj_0-D_0\nabla^2\vec{A}
              \right.
             \nonumber\\
       &&\mbox{}\left.+g_0\nabla\psi_0(r_0-\nabla^2)\psi_0
         -g_0\nabla\psi_0h+\vec{\zeta}\right].
\label{samlan}
\end{eqnarray}
Here, $g_0$ denotes the strength of the reversible mode-coup\-ling 
terms, and $\eta$ and $\vec{\zeta}$ represent fluctuating forces with 
zero mean, $\langle \eta({\bf x},t) \rangle = 0$, 
$\langle \zeta^{\alpha}({\bf x},t) \rangle = 0$. The couplings 
$\lambda_0$ and $D_0$ are, respectively, the thermal conductivity and 
shear viscosity (in appropriate units). ${\cal T}[\ldots]$ denotes a 
projection operator which selects the transverse part of the vector 
in brackets; in Fourier space ${\cal T}^{\alpha\beta}(\vec{k})= 
\delta^{\alpha\beta}-k^\alpha k^\beta/k^2$. The above Langevin-type 
equations of motion are invariant under Galilean transformations, a 
symmetry which can be utilized in the renormalization procedure (see 
below and the appendix).

In order to fully characterize the dynamics, we furthermore need to 
specify the correlations of the stochastic forces. Given that both the 
order parameter and the transverse current are conserved quantities, the 
strength of the random forces has to vanish at zero momentum. We assume 
that these random fields are Gaussian correlated and we write, in general, 
the second moment of the distributions in Fourier space as:
\begin{equation}
        \langle \eta(\vec{k},\nu) \eta(\vec{k}',\nu') 
        \rangle = 2 {\widetilde \lambda}_0 (\vec{k})\, 
        \delta(\vec{k} + \vec{k}') \,
                        \delta( \nu + \nu') 
 \label{sopnoi}
\end{equation}
and
\begin{eqnarray}
        \langle \zeta^{\alpha}(\vec{k},\nu) 
                \zeta^{\beta}(\vec{k}',\nu') \rangle &=&
        2 {\widetilde D}_0(\vec{k})\,
                \delta(\vec{k} +\vec{k}') \,
                \delta( \nu + \nu') \nonumber \\
        &&\mbox{}\times \left(\delta^{\alpha\beta}-
		\frac{k^\alpha k^\beta}{k^2}\right) ,
 \label{samnoi}
\end{eqnarray}
where the transverse projector in (\ref{samnoi}) insures that the 
random force is in the transverse directions only. For the equilibrium 
model H, the functions $\widetilde{\lambda}_0(\vec{k})$ and 
$\widetilde{D}_0(\vec{k})$ are equal to 
$\widetilde{\lambda}_0(\vec{k})=\lambda_0\,k_B T\,k^2$,
$\widetilde{D}_0(\vec{k})=D_0\,k_B T\,k^2$, i.e., the noise correlators 
satisfy the Einstein relations and the conservation conditions 
$\widetilde{\lambda}_0(\vec{0})=\widetilde{D}_0(\vec{0})=0$.

As stated above, we will consider two choices for the noise correlators
which do not satisfy the Einstein conditions and therefore describe
{\em non-equilibrium} versions of model H. Both these choices can be 
viewed as generic perturbations from the equilibrium situation. In the 
first case, we take:
\begin{equation}
\widetilde{\lambda}_0(\vec{k})=\widetilde{\lambda}_0\,k^2
\label{isomodelH1}
\end{equation}
and
\begin{equation}
\widetilde{D}_0(\vec{k})=\widetilde{D}_0\,k^2,
\label{isomodelH2}
\end{equation}
with $\widetilde{\lambda}_0/\lambda_0\neq\widetilde{D}_0/D_0$. Since 
the equilibrium model H corresponds to the case in which 
$\widetilde{\lambda}_0/\lambda_0=\widetilde{D}_0/D_0=k_B T$, one may
interpret Eqs.~(\ref{isomodelH1}) and (\ref{isomodelH2}) as describing 
a situation in which the two dynamical variables, the order parameter 
and the transverse current, are effectively placed in contact with
heat baths at {\em different} temperatures. 

In the second case, we take more generally:
\begin{equation}
\widetilde{\lambda}_0(\vec{k})=\widetilde{\lambda}^{\parallel}_0\,
k_{\parallel}^2+\widetilde{\lambda}^{\perp}_0\,k_{\perp}^2
\label{animodelH1}
\end{equation}
and
\begin{equation}
\widetilde{D}_0(\vec{k})=\widetilde{D}^{\parallel}_0\,k_{\parallel}^2
+\widetilde{D}^{\perp}_0\,k_{\perp}^2 \ .
\label{animodelH2}
\end{equation}
This choice of the noise correlators corresponds to a situation
where two spatial sectors of dimensions $d_{\parallel}$ and $d_{\perp}$
are placed at different temperatures, i.e., where real space isotropy
is broken \cite{orderparani}. Without loss of generality, we set
the effective noise temperature higher in the parallel subspace.

Once the noise distribution is specified, one can represent the 
Langevin equations (\ref{hoplan}) and (\ref{samlan}), with 
(\ref{sopnoi}) and (\ref{samnoi}), as a dynamic field theory, following 
standard procedures \cite{dynfun,uwezol}. This results in a probability 
distribution for the dynamic fields $\psi_0$
and $\vecj_0$,
\begin{eqnarray}
        P[\{ \psi_0, \vecj_0 \}] &\propto&
        \int {\cal D}[\{ i {\widetilde \psi}_0\}] \int 
        {\cal D}[\{ i \,\widetilde{\vecj}_0 \}] 
        \times \nonumber \\
        &&\times \, e^{J[\{ {\widetilde \psi}_0\}, \{ \psi_0 \},
        \{ \widetilde{\vecj}_0 \}, 
        \{ \vecj_0 \}]} \ ,
 \label{genfun}
\end{eqnarray}
with the statistical weight given by the Janssen-De~Dominicis functional
$J = J_{\rm har} + J_{\rm rel} + J_{\rm mc}+ J_{\rm sc}$, which we divide 
into an harmonic part $J_{\rm har}$, which one can integrate exactly; the
purely relaxational (of static origin) and reversible dynamic non-linear 
terms $J_{\rm rel}$ and $J_{\rm mc}$, which are to be expanded in a power
series, giving rise to the perturbation series in terms of Feynman 
diagrams; and a term $J_{\rm sc}$ which depends on the external sources.

The harmonic part, in terms of the original dynamic fields 
$\psi_0({\bf k},\nu)$, $\vecj_0({\bf k},\nu)$ and the 
auxiliary fields ${\widetilde \psi}_0({\bf k},\nu)$, 
$\widetilde{\vecj}_0({\bf k},\nu)$ (we use, for
convenience, the Fourier space representation) reads
\begin{eqnarray}
\label{harmfun}
        &&J_{\rm har}[\{{\widetilde \psi}_0\}, \{ \psi_0 \},
        \{ \widetilde{\vecj}_0 \}, 
        \{ \vecj_0 \}] \\
        &&= \int \! \frac{d^dk_1}{(2\pi)^d} \, 
         \frac{d\nu_1}{2\pi} \,\Biggl[
          {\widetilde \lambda}_0(\vec{k}_1) 
 {\widetilde \psi}_0(-\vec{k}_1,-\nu_1) 
\,{\widetilde \psi}_0(\vec{k}_1,\nu_1)
 \nonumber\\
  &&\mbox{}+\sum_{\alpha,\beta}\,{\widetilde D}_0(\vec{k}_1)\,
 \widetilde{\scj}_0^\alpha(-\vec{k}_1,-\nu_1)
 {\cal T}^{\alpha\beta}(\vec{k}_1)\,
 \widetilde{\scj}_0^\beta (\vec{k}_1,\nu_1)\nonumber \\
 &&\mbox{}-{\widetilde \psi}_0(-\vec{k}_1,-\nu_1)
        \left[-i\nu_1 + \lambda_0 k_1^2\left( r_0 +k_1^2 \right) 
                \right] \psi_0(\vec{k}_1,\nu_1)\nonumber \\ 
        &&\mbox{} - \sum_{\alpha,\beta}
        \widetilde{\scj}_0^{\alpha}(-\vec{k}_1,-\nu_1)\!
        \left( -i\nu_1 +D_0 k_1^2 \right)\!
        {\cal T}^{\alpha\beta}(\vec{k}_1)\,
                \scj_0^{\beta}(\vec{k}_1,\nu_1) \Biggr] \ ,
\nonumber 
\end{eqnarray}
where the projector ${\cal T}^{\alpha\beta}(\vec{k}_1)$ insures that
only the transverse components of the fields $\widetilde{\scj}_0^\alpha$,
$\scj_0^{\beta}$ contribute to the action. The longitudinal component,
as mentioned before, does not represent independent fluctuations, and its
contributions (though formally infinite) can be factored out of the 
functional integral.

The static non-linearity leads, in turn, to the relaxation vertex
\begin{eqnarray}
\label{relfun}
        &&J_{\rm rel}[\{ {\widetilde \psi}_0 \} , \{ \psi_0 \}] \\
        &&\mbox{}= 
        - {\lambda_0 u_0 \over 6} \int \! \frac{d^dk_1}{(2\pi)^d}
        \,\frac{d^dk_2}{(2\pi)^d} \, \frac{d^dk_3}{(2\pi)^d}
        \, \frac{d\nu_1}{2\pi}
        \,\frac{d\nu_2}{2\pi}
        \, \frac{d\nu_3}{2\pi}\nonumber\\
        &&\mbox{}\times k^2
        {\widetilde \psi}_0(-\vec{k},-\nu)\,
        \psi_0(\vec{k}_1,\nu_1)\,
        \psi_0(\vec{k}_2,\nu_2)\, \psi_0(\vec{k}_3,\nu_3) ,
       \nonumber
 \end{eqnarray}
where $(\vec{k},\nu)=(\vec{k}_1+\vec{k}_2+\vec{k}_3,\nu_1+\nu_2+\nu_3)$.
The purely dynamic couplings generate the mode-cou\-pling vertices
\begin{eqnarray}
 \label{mctfun}
        &&J_{\rm mc}[\{{\widetilde \psi}_0\}, \{ \psi_0 \},
        \{ \widetilde{\vecj}_0 \}, 
        \{ \vecj_0 \}]
        \nonumber\\
        &&=  -ig_0\int \! \frac{d^dk_1}{(2\pi)^d}
        \,\frac{d^dk_2}{(2\pi)^d}
        \, \frac{d\nu_1}{2\pi}
        \,\frac{d\nu_2}{2\pi} \sum_{\alpha,\beta}
        \nonumber\\ 
        &&\mbox{}\left[\,
           k_1^{\alpha}\,
        {\cal T}^{\alpha\beta}(\vec{k}_2)
        \,\widetilde{\psi}(-\vec{k},-\nu)
        \psi(\vec{k}_1,\nu_1)\,
        \scj^{\beta}(\vec{k}_2,\nu_2)\right.\nonumber\\
        &&\mbox{}
        -\frac{1}{2}\,
        (k_1^{\alpha}k_2^2+k_2^{\alpha}k_1^2)
        \,{\cal T}^{\alpha\beta}(\vec{k})
        \nonumber\\
        &&\mbox{}\left.\times \,
        \widetilde{\scj}^{\beta}(-\vec{k},-\nu)\,
        \psi(\vec{k}_1,\nu_1) 
        \psi(\vec{k}_2,\nu_2)\,\right]
\end{eqnarray}
with $(\vec{k},\nu)=(\vec{k}_1+\vec{k}_2,\nu_1+\nu_2)$ here.

Finally, the source terms give rise to the following
contribution
\begin{eqnarray}
\label{scfun}
        &&J_{\rm sc}[\{{\widetilde \psi}_0\}, \{ \psi_0 \},
        \{ \widetilde{\vecj}_0 \}, 
        \{ \vecj_0 \}]
        \\
        &&=
        \int \! \frac{d^dk_1}{(2\pi)^d} 
         \frac{d\nu_1}{2\pi}
         \biggl[\lambda_0 k_1^2\,
         \widetilde{\psi}(-\vec{k}_1,-\nu_1)\,h(\vec{k}_1,\nu_1) \nonumber\\
        &&\mbox{}+ \sum_{\alpha,\beta}
        D_0 k_1^2 \,\widetilde{\scj}^{\beta}(-\vec{k}_1,-\nu_1)
           \,{\cal T}^{\alpha\beta}(\vec{k}_1)\,
           A^{\beta}(\vec{k}_1,\nu_1)\biggr]
         \nonumber\\ 
        &&\mbox{}+ig_0\int \! \frac{d^dk_1}{(2\pi)^d}
        \,\frac{d^dk_2}{(2\pi)^d}
        \, \frac{d\nu_1}{2\pi}
        \,\frac{d\nu_2}{2\pi} \sum_{\alpha,\beta}
        \nonumber\\
        &&\mbox{} \left[\,
           k_1^{\alpha}\,
        {\cal T}^{\alpha\beta}(\vec{k}_2)
        \,\widetilde{\psi}(-\vec{k},-\nu)
        \psi(\vec{k}_1,\nu_1)\,
        A^{\beta}(\vec{k}_2,\nu_2)\right.
        \nonumber\\
        &&\mbox{} \left. \,k_2^{\alpha}
        \,{\cal T}^{\alpha\beta}(\vec{k})\,
        \widetilde{\scj}^{\beta}(-\vec{k},-\nu)\,
        h(\vec{k}_1,\nu_1) 
        \psi(\vec{k}_2,\nu_2)\,\right]\,\nonumber
\end{eqnarray}
with $(\vec{k},\nu)=(\vec{k}_1+\vec{k}_2,
\nu_1+\nu_2)$ in the second integral. The contribution of the sources 
is simply included for convenience. If one wishes to relate the dynamic 
susceptibilities to the Green's functions, one can simply differentiate
the dynamic functional with respect to $h(\vec{x},t)$, $\vec{A}(\vec{x},t)$
and then set these sources equal to zero. The expression of the dynamic
functional in Fourier space (\ref{harmfun}) through (\ref{mctfun}),
although more cumbersome in notation, is directly applicable to the
perturbation expansion. 
 
As usual, the harmonic part (\ref{harmfun}) defines the propagators of the
field theory, while the perturbation expansion is performed in terms of the
non-linear vertices (\ref{relfun}) and (\ref{mctfun}).
Notice that the existence of the reversible forces (\ref{mctfun}) does not
show up in dynamic mean-field theory (van Hove theory), which in field-theory
language is based on the harmonic action (\ref{harmfun}) only.
We shall see that while the choice (\ref{isomodelH1}), (\ref{isomodelH2})
for the functions $\widetilde{\lambda}_0(\vec{k})$ and 
$\widetilde{D}_0(\vec{k})$ yields a perfectly consistent field theory, 
the choice (\ref{animodelH1}), (\ref{animodelH2}) predicts a correction 
to the critical temperature which is {\em anisotropic}. One therefore 
needs to modify the theory accordingly in order to obtain a consistent 
description of the model. We shall address this issue in section 
\ref{anisoH}. Here it suffices to say that once we have modified the 
Langevin equations describing the dynamics, we can also treat the 
resulting model by the means described above.

Once the dynamic functional is obtained, the perturbation expansion for 
all possible correlation functions of the dynamic and auxiliary fields, 
as well as for the associated vertex functions, is given by the 
one-particle irreducible Feynman diagrams. A straightforward scaling 
analysis yields that the upper critical dimension of the isotropic model 
H is $d_c = 4$ for both the relaxational {\em and} the mode-coupling 
vertices. For the anisotropic, effective two-temperature model H, 
however, the upper critical dimension is reduced to $d_c = 
4-d_{\parallel}$, as will be seen below. For $d \leq d_c$, the 
perturbation theory will be infrared-singular, and non-trivial critical 
exponents ensue, while for $d \geq d_c$ the perturbation theory 
contains ultraviolet divergences. In order to renormalize the field 
theory in the ultraviolet, it suffices to render all the non-vanishing 
two-, three-, and four-point functions finite by introducing 
multiplicative renormalization constants, following an additive 
renormalization corresponding to a fluctuation-induced shift $r_{0c}$ 
of the critical temperature. This is achie\-ved by demanding the 
renormalized vertex functions, or appropriate momentum and frequency 
derivatives thereof, to be finite when the fluctuation integrals are 
taken at a conveniently chosen normalization point, well outside the 
singularities of the infrared regime.

We shall employ the dimensional regularization scheme in order to 
compute the emerging momentum integrals, and choose the renormalized 
mass $\tau = 1$ as our normalization point, or, sufficient to one-loop 
order, $\tau_0 = r_0 - r_{0c} = \mu^2$. Notice that $\mu$ defines an 
intrinsic momentum scale of the renormalized theory. From the 
renormalization constants ($Z$ factors) that render the field theory
finite in the ultraviolet (UV), one may then derive the renormalization
group (RG) flow functions which enter the Gell-Mann--Low equation. This 
partial differential equation describes how correlation functions change 
under scale transformations. In the vicinity of an RG fixed point, the 
theory becomes scale-invariant, and the information previously gained 
about the UV behavior can thus be employed to access the physically 
interesting power laws governing the infrared (IR) regime at the 
critical point ($\tau \propto T - T_c \to 0$) for long wavelengths 
(wave vector ${\vec q} \to 0$) and low frequencies ($\omega \to 0$).

\section{Renormalization group analysis of the isotropic model H}
\label{isoH}
\subsection{Vertex and response function renormalization}
\label{verresp}

The UV-divergent two-, three-, and four-point vertex functions or their
derivatives that require multiplicative renormalization are
$\partial_\omega \Gamma_{0 \, \widetilde{\psi} \psi}({\bf q},\omega)$,
$\partial_{q^2} \Gamma_{0 \, \widetilde{\psi} \psi}({\bf q},\omega)$
together with
$\partial_{q^4}\Gamma_{0 \, \widetilde{\psi} \psi}({\bf q},\omega)$
and $\partial_{q^2} \Gamma_{0 \, \widetilde{\psi} \widetilde{\psi}}(
{\bf q},\omega)$; the functions involving $\vecj$,
$\widetilde{\vecj}$,
$\partial_\omega \Gamma_{0 \, \widetilde{\scj} \,\scj}({\bf q},\omega)$,
$\partial_{q^2} \Gamma_{0 \, \widetilde {\scj} \,\scj}({\bf q},\omega)$
and $\partial_{q^2} \Gamma_{0 \, \widetilde{\scj}\, \widetilde{\scj}}(
{\bf q},\omega)$; the three-point vertices 
$\partial_{q^\alpha} 
\Gamma_{0 \, \widetilde{\psi}\psi\,\scj^{\alpha}}(\vec{-q-p},
-\omega;\vec{q},\omega;\vec{p},0)$ and
$\partial_{(q^\alpha p^2+p^\alpha q^2)}
\Gamma_{0 \widetilde{\scj}^{\alpha}\psi\psi}\!(\vec{-q-p},
-\omega;\vec{q},\omega;\vec{p},0)$; and finally the rela\-xation vertex
$\partial_{q^2} 
\Gamma_{0 \widetilde{\psi} \psi\psi\psi}\!(-{\bf q},\!-\omega;
{{\bf q} \over 3},{\omega \over 3}; {{\bf q} \over 3},{\omega \over 3};
{{\bf q} \over 3},{\omega \over 3})$.
On the other hand, we have four fluctuating fields 
($\widetilde{\scj}_0^{\alpha}$, $\scj_0^{\alpha}$, 
$\widetilde{\psi}_0$, $\psi_0$) and the seven parameters 
${\widetilde D}_0$, $D_0$, ${\widetilde \lambda}_0$, $\lambda_0$, $\tau_0$, 
$g_0$ and $u_0$ available; this leaves us at liberty to choose one of the
renormalization constants in a convenient manner.
In addition to these parameters and since
detailed balance does not hold for this non-equilibrium model,
one also needs to consider the renormalization of 
the dynamic susceptibility of the order parameter,
in order to determine the susceptibility exponent $\eta$.

Starting with the two-point functions 
$\Gamma_{0 \, \widetilde{\psi} \psi}({\bf q},\omega)$ and 
$\Gamma_{0 \, \widetilde{\scj}\, \scj}({\bf q},\omega)$ 
for the conserved order parameter and transverse current fluctuations, 
respectively, we immediately note that as a 
consequence of the momentum dependence of the mode-coupling 
vertices, one has for these two vertex functions
\begin{equation}
        {\partial \over \partial (i \omega)} 
        \Gamma_{0 \,\widetilde{\psi} \psi\,
(\widetilde{\scj}\, \scj)}({\bf q} = {\bf 0},\omega) \equiv 1
 \label{amverf}
\end{equation}
to {\em all orders} in perturbation theory. 
Upon defining renormalized fields according to
\begin{eqnarray}
        &&\widetilde{\psi}= Z_{\widetilde{\psi}}^{1/2} 
                \widetilde{\psi}_0 \; , \quad 
        \psi= Z_{\psi}^{1/2} \psi_0\ ,
 \label{conren} \\
        &&\widetilde{\scj}^\alpha = Z_{\widetilde{\scj}}^{1/2} 
                \widetilde{\scj}_0^\alpha \; , \quad 
        \scj^\alpha = Z_{\scj}^{1/2} \scj_0^\alpha \ ,
 \label{fldren}
\end{eqnarray}
which imply that $\Gamma_{\widetilde{\psi} \psi} 
= (Z_{\widetilde{\psi}} Z_{\psi})^{-1/2} 
\Gamma_{0 \, \widetilde{\psi} \psi}$
and that $\Gamma_{\widetilde{\scj}\, \scj} 
= (Z_{\widetilde{\scj}} Z_{\scj})^{-1/2} 
\Gamma_{0 \, \widetilde{\scj}\, \scj}$, 
we thus obtain the exact relations
\begin{eqnarray}
Z_{\widetilde{\psi}} \,Z_\psi &\equiv& 1 \ ,
\label{zmtrel1}
\\
Z_{\widetilde{\scj}} \,Z_{\scj} &\equiv& 1 \ .
 \label{zmtrel2}
\end{eqnarray}
At this point we utilize our freedom of choice \cite{eqth} to set
\begin{equation}
        Z_{\widetilde{\scj}} \equiv Z_{\scj} \equiv 1 \ .
 \label{zmtcho}
\end{equation}
The multiplicative renormalization factors for the coupling constants
are defined through
\begin{eqnarray}
\lambda&=&Z_{\lambda}\lambda_0 \ ,
\label{RGlambda}\\
\widetilde{\lambda}&=&Z_{\widetilde{\lambda}}\widetilde{\lambda}_0 \ ,
\label{RGlambdatilde}\\
D&=&Z_{D}\,D_0\,\mu^{-2} \ ,
\label{RGD}\\
\widetilde{D}&=&Z_{\widetilde{D}}\widetilde{D}_0\mu^{-2} \ ,
\label{RGDtilde}\\
\tau&=&Z_{\tau}\,\tau_0\,\mu^{-2}\;\;\mbox{with}\;\;\tau_0=r_0-r_{0c} \ ,
\label{RGtau}\\
u&=&Z_{u}\,u_0\,A_d\,\mu^{d-4} \ ,
\label{RGu}\\
g&=&Z_g^{1/2}\,g_0\,A_d^{1/2}\,\mu^{d/2-3} \ ,
\label{RGg}
\end{eqnarray}
where the $\mu$ factors have been introduced to render the renormalized
couplings dimensionless. The geometric numerical factor 
$A_d=\Gamma(3-d/2) / 2^{d-1} \pi^{d/2}$ has been absorbed into $u$ and $g$
since it appears in the explicit integrals of the subsequent 
perturbational analysis. Finally, we still have  at our disposal a 
symmetry of the theory which is valid even in a non-equilibrium situation, 
namely Galilean invariance (see Appendix \ref{appA}). This symmetry 
imposes the exact condition that
\begin{equation}
        Z_{g} \, Z_{\scj} \equiv 1 \ .
 \label{galZ}
\end{equation}
From this relation and Eq.~(\ref{zmtcho}) it follows that the dynamical 
vertex does not renormalize, i.e. $Z_{g}\equiv 1$.

In order to discuss the RG flow diagram, it is convenient
to introduce the following rescaled coupling constants
\begin{eqnarray}
\widetilde{u}_0&=&\frac{\widetilde{\lambda}_0}{\lambda_0}\,u_0 \ ,
\label{tildeu0}
\\
\widetilde{f}_0&=&\frac{\widetilde{\lambda}_0}{\lambda_0}
\,\frac{g_0^2}{\lambda_0 D_0} \ ,
\label{tildef0}
\end{eqnarray}
and the ratio between the noise temperature of the order parameter
and the noise temperature of the mass-energy current, respectively 
\cite{conT0}
\begin{equation}
T_0=\frac{\widetilde{\lambda}_0}{\lambda_0}\,\frac{D_0}{\widetilde{D}_0}\ ,
\label{eqT0}
\end{equation}
with the renormalized versions of these coupling constants being
defined as
\begin{eqnarray}
\widetilde{u}&=&\frac{Z_{\widetilde{\lambda}}Z_u}{Z_\lambda}\,
\widetilde{u}_0\,A_d\,\mu^{d-4}\,,
\label{tildeu}
\\
\widetilde{f}&=&\frac{Z_{\widetilde{\lambda}}}{Z_\lambda^2 Z_D}\,
\widetilde{f}_0\,A_d\,\mu^{d-4}\,,
\label{tildef}\\
T&=&\frac{Z_{\widetilde{\lambda}}Z_D}{Z_{\lambda}Z_{\widetilde{D}}}\,T_0\, ,
\label{eqT}
\end{eqnarray}
whence the upper critical dimension of the dynamical vertex 
$\widetilde{f}$ comes out to be the same as that of the static
vertex $\widetilde{u}$, i.e. $d_c=4$. Also, in equilibrium, 
$Z_{\widetilde{\lambda}}= Z_{\lambda}$, $Z_{\widetilde{D}}=Z_D$, 
wherefrom we conclude that $T=T_0=1$.

The analysis is now carried through by considering all one-loop
Feynman diagrams which contribute to the different vertex functions
listed above. The renormalization factors are determined
in such a way that the renormalized vertex functions depend on
the renormalized couplings in the same way as the zero-loop
vertex functions depend on the bare couplings at a normalization
point, chosen here to be $(\vec{q},\omega)=(\vec{0},0)$ and $\tau=1$
(to one-loop order, $\tau_0=\mu^2$), well outside the IR region.
The introduction of this normalization point (NP) renders the renormalized 
couplings dependent on the momentum scale $\mu$, as pointed out above.   

From $\partial_{q^2} \Gamma_{ \, \widetilde{\psi} 
\psi}({\bf q},\omega)|_{\rm NP}$ and $\partial_{q^4}\Gamma_{ \, 
\widetilde{\psi} \psi}({\bf q},\omega)|_{\rm NP}$,
one obtains the following results for the renormalization
factors $Z_{\lambda}$ and $Z_{\tau}$ \cite{IRfint},
\begin{eqnarray}
Z_{\lambda}\,Z_{\tau}&=&1+\frac{2(d-1)}{(d-2)d}\,\widetilde{f}_0\,
\frac{A_d}{\epsilon}\mu^{-\epsilon}\nonumber\\
&&\mbox{}-\frac{\widetilde{u}_0}{d-2}\,
\,\frac{A_d}{\epsilon}\mu^{-\epsilon}\,,
\label{ZlZt}\\
Z_{\lambda}&=&1+\frac{2(d-1)}{(d-2)d}\,\widetilde{f}_0\,
\frac{A_d}{\epsilon}\mu^{-\epsilon}\,,
\label{Zl}
\end{eqnarray}
where we have used the value of $r_{0c}$ obtained from the
condition of divergence of the static susceptibility (see below)
and where $\epsilon=4-d$. From these equations one can obtain the
value of $Z_{\tau}$, namely
\begin{equation}
Z_{\tau}=1-\frac{\widetilde{u}_0}{d-2}\,
\frac{A_d}{\epsilon}\mu^{-\epsilon}\,.
\label{Zt}
\end{equation}
Applying the normalization condition to the derivative 
$\partial_{q^2} \Gamma_{ \, \widetilde{\psi} 
\widetilde{\psi}}({\bf q},\omega)|_{\rm NP}$, one obtains
the following expression for $Z_{\widetilde{\lambda}}/Z_{\psi}$
\begin{equation}
\frac{Z_{\widetilde{\lambda}}}{Z_\psi}=1+\frac{2(d-1)}{(d-2)d}\,
\frac{\widetilde{f}_0}{T_0}\,
\frac{A_d}{\epsilon}\,\mu^{-\epsilon}\,.
\label{ZtlZpsi}
\end{equation}
The normalization conditions for the derivatives of the vertex functions
$\partial_{q^2} \Gamma_{ \, \widetilde {\scj} 
\,\scj}({\bf q},\omega)|_{\rm NP}$ and $\partial_{q^2} \Gamma_{ \, 
\widetilde{\scj}\, \widetilde{\scj}}({\bf q},\omega)|_{\rm NP}$ in turn 
yield the following expressions for $Z_D$ and $Z_{\widetilde{D}}$,
\begin{eqnarray}
Z_{D}&=&1+\frac{\widetilde{f}_0}{4(d+2)}\,\frac{A_d}{\epsilon}\,
\mu^{-\epsilon}\,,
\label{ZD}\\
Z_{\widetilde{D}}&=&1+\frac{1}{4(d+2)}\,\widetilde{f}_0\,T_0\,
\frac{A_d}{\epsilon}\,\mu^{-\epsilon}\,,
\label{ZtD}
\end{eqnarray}
which become equal in equilibrium, $T_0=1$, as they should. From the 
normalization condition for the three point vertex $\partial_{q^\alpha} 
\Gamma_{\,\widetilde{\psi}\psi\,\scj^{\alpha}}(\vec{-q-p},
-\omega;\vec{q},\omega;\vec{p},0)|_{\rm NP}$, one simply obtains,
to one-loop order, that $Z_g=1$, which confirms the result, valid to 
all orders, obtained from the condition of Galilean invariance. 

On the other hand, from the normalization condition for 
$\partial_{(q^\alpha p^2+p^\alpha q^2)}
\Gamma_{ \, \widetilde{\scj}^{\alpha}\psi\psi}(\vec{-q-p},
-\omega;\vec{q},\omega;\vec{p},0)\mid_{NP}$, one
obtains the following expression for $Z_{\psi}$
\begin{equation}
Z_{\psi}=1+\frac{4\widetilde{f}_0}{(d-2)d(d+2)}\left(
1-\frac{1}{T_0}\right)\,
\frac{A_d}{\epsilon}\,\mu^{-\epsilon}\,.
\label{Zpsi}
\end{equation}
Note that in equilibrium this expression is identical to 1, i.e., 
there are no corrections to $\eta = 0$ to one-loop order.
Eq.~(\ref{Zpsi}) can be used in (\ref{ZtlZpsi}) to find
$Z_{\widetilde{\lambda}}$,
\begin{eqnarray}
\label{Ztl}
Z_{\widetilde{\lambda}}&=&1+\frac{2(d-1)}{(d-2)d}\,
\frac{\widetilde{f}_0}{T_0}\,\frac{A_d}{\epsilon}\,\mu^{-\epsilon}
\\
&&\mbox{}+\frac{4}{(d-2)d(d+2)}\,\widetilde{f}_0\,
\left(1-\frac{1}{T_0}\right)\,
\frac{A_d}{\epsilon}\,\mu^{-\epsilon}\,,\nonumber
\end{eqnarray}
which again reduces to (\ref{Zl}) in equilibrium.

Finally, the normalization condition for the relaxation vertex
$\partial_{q^2} 
\Gamma_{ \, \widetilde{\psi} \psi\psi\psi}(-{\bf q},-\omega;
{{\bf q} \over 3},{\omega \over 3}; {{\bf q} \over 3},{\omega \over 3};
{{\bf q} \over 3},{\omega \over 3})|_{\rm NP}$ yields the following
expression for the product $Z_{\lambda}\,Z_u\,Z_{\psi}$,
\begin{eqnarray}
Z_{\lambda}\,Z_u\,Z_{\psi}&=&1-\frac{3}{2}\,\widetilde{u}_0\,
\frac{A_d}{\epsilon}\,\mu^{-\epsilon}\nonumber\\
&&\mbox{}+\frac{d-1}{d}\,\widetilde{f}_0\,
\frac{A_d}{\epsilon}\,\mu^{-\epsilon}\,,
\label{ZlZuZpsi}
\end{eqnarray}
from which one obtains, using (\ref{Zl}) and (\ref{Zpsi}),
the following result for $Z_u$,
\begin{eqnarray}
&&Z_u=1-\frac{3}{2}\,\widetilde{u}_0\,
\frac{A_d}{\epsilon}\,\mu^{-\epsilon}\nonumber\\
&&\ \mbox{}-\frac{4}{(d-2)d(d+2)}\,\widetilde{f}_0\,
\left(1-\frac{1}{T_0}\right)\,
\frac{A_d}{\epsilon}\,\mu^{-\epsilon}\,,
\label{Zu}
\end{eqnarray}
which is independent of the dynamic coupling $\widetilde{f}_0$ 
in equilibrium, as it should be.
 
The above results for the Z factors are sufficient to provide the
beta functions which determine the fixed points of the theory. 
However, one still needs to consider the independent renormalizations
needed for the susceptibilities, since one does not have the usual
constraints imposed by detailed balance. To define the 
``static'' limit of the intrinsically dynamic model 
under consideration, we compute the response functions for the 
order parameter and for the transverse current, and then take the limit
$\omega \to 0$ there. In order to obtain these functions, one simply
needs to take the derivative of the functional $J_{sc}$, equation
(\ref{scfun}), with respect 
to the external sources $h(\vec{x},t)$ and $\vec{A}(\vec{x},t)$
and then set these sources equal to zero. Some formal manipulations 
using the properties of the
vertex functions yield
\begin{eqnarray}
\chi_0({\bf q},\omega) &=& 
        \Gamma_{0\, \widetilde{\psi}\,\psi}(-{\bf q},-\omega)^{-1} 
\nonumber \\
        &&\times \left[ \lambda_0\,q^2 + g_0 \,
        \Gamma_{0\, \widetilde{\psi}\,[\widetilde{\scj}\nabla\psi]}
(-{\bf q},-\omega) 
        \right] \ ,
 \label{opvsus}\\
        X_0({\bf q},\omega) &=& 
        \Gamma_{0\, \widetilde{\scj}\,\scj}(-{\bf q},-\omega)^{-1} 
\nonumber \\
        &&\times \left[ D_0 q^2 - g_0 \,
        \Gamma_{0\, \widetilde{\scj}\,[\widetilde{\psi}\nabla\psi]}(-{\bf q},
-\omega) 
        \right] ,
 \label{amvsus}
\end{eqnarray}
respectively \cite{dynfun,uwezol,factorPout}.
Note that composite-operator vertex functions enter these expressions, which
in general implies that new renormalization constants are required to remove 
the UV singularities of the response functions (equivalently, one may utilize
the $Z$ factors obtained from the multiplicative renormalization of the vertex
functions plus appropriate additive renormalizations \cite{dynfun,uwezol}).
Yet one may show to {\em all orders} in perturbation theory \cite{notequadr}
that
\begin{equation}
\Gamma_{0\, \widetilde{\scj}\scj}({\bf q},\omega) = i \omega + D_0 q^2 - g_0 \,
\Gamma_{0\, \widetilde{\scj}\,[\widetilde{\psi}\nabla\psi]}({\bf q}, \omega)\, 
, \label{susidt}
\end{equation}
and consequently
\begin{equation}
        X_0({\bf q},\omega=0) \equiv 1 \ , 
 \label{stamsu}
\end{equation}
which means that there is no additional renormalization required here.
On the other hand, the static limit of the order parameter susceptibility is in
fact singular, which leads us to define the corresponding renormalized response
function via
\begin{equation}
        \chi({\bf q},\omega) = Z \, \chi_0({\bf q},\omega) \ .
 \label{rensus} 
\end{equation}
The new renormalization constant $Z$ is determined by demanding that 
$\partial_{q^2} \chi^{-1}({\bf q},\omega)\vert_{\rm NP}$ be UV-finite. 
In this case, to one-loop order we obtain the simple result $Z=1$, i.e., 
there are no corrections to this order of perturbation theory. 
Note that in equilibrium, one has $Z_{\psi}=Z$ to all orders as a 
consequence of detailed balance \cite{care1}.

The fluctuation-induced $T_c$ shift is determined from the criticality
condition $\chi_0^{-1}({\bf q} = {\bf 0},\omega = 0)= 0$ at $r_0 = r_{0c}$
($\tau_0 = 0$) with the result
\begin{eqnarray}
\label{r0c}
        &&r_{0c} = - {1 \over 2} \, {\widetilde u}_0 
                \int_k \! {1 \over r_{0c} + k^2} 
 \label{tcsint} \\
        &&\mbox{}+\frac{d-1}{d}\,\frac{D_0}{\lambda_0}\,\widetilde{f}_0\,
 \left( 1 - \frac{1}{T_0}\right)\,
                \int_k \! {1 \over k^2(r_{0c}+D_0/\lambda_0 
+ k^2)} \, , \nonumber
\end{eqnarray}
which determines $r_{0c}$ implicitly.
The momentum integrals in (\ref{r0c}) should be evaluated with a {\em finite} 
upper cutoff, which underlines the non-universality of the $T_c$ shift, i.e.,
its dependence on short-distance properties. This expression is however
sufficient, in its present form, to provide for the additive renormalization
(mass renormalization) necessary to make 
$\Gamma_{\widetilde{\psi} \psi}({\bf q},\omega)$ 
finite and we have implicitly
used it to obtain the results (\ref{ZlZt}) and (\ref{Zl}).

\subsection{Discussion of the RG flow equations}
 \label{sssflo}

\subsubsection{RG equations for the vertex and response functions}
\label{isorgver}

By means of the above renormalization constants, we can now write down the RG
(Gell-Mann--Low) flow equations for the vertex functions and the dynamic
susceptibilities. 
The latter connect the asymptotic theory, where the IR singularities become 
manifest, with a region in parameter space where the loop integrals are finite 
and ordinary `naive' perturbation expansion is applicable, and follow from 
the simple observation that the `bare' vertex functions do not depend on the 
renormalization scale $\mu$,
\begin{equation}
        \mu {d \over d\mu} \bigg\vert_0 \Gamma_{0 \, \widetilde{\scj}^k 
        \widetilde{\psi}^r \scj^l \psi^s}(\{ {\bf q},\omega \} 
; \{ a_0 \}) = 0 \ ,
 \label{rngreq}
\end{equation}
with $\{ a_0 \} =\lambda_0$, $\widetilde{\lambda}_0$,
$D_0$, ${\widetilde D}_0$, 
$\tau_0$, $u_0$ and $g_0$.
Replacing the bare parameters and fields in Eq.~(\ref{rngreq}) with the
renormalized ones, we thus find the following partial differential equations
for the renormalized vertex functions,
\begin{eqnarray}
        &&\left[ \mu {\partial \over \partial \mu} 
        + \sum_{\{ a \}} \zeta_a\,
        a\,{\partial \over \partial a} 
        + {r \over 2} \, \zeta_{\widetilde{\psi}} 
        + {s \over 2} \, \zeta_\psi \right] \nonumber \\ 
        &&\quad \times \Gamma_{\widetilde{\scj}^k \widetilde{\psi}^r \scj^l 
\psi^s} 
        \left( \mu, \{ {\bf q},\omega \};
                \{ a \} \right) = 0 \ .
 \label{calsym}
\end{eqnarray}
Here, we have introduced Wilson's flow functions
\begin{eqnarray}
        \zeta_\psi&=& \mu {\partial \over \partial \mu} 
                \bigg\vert_0 \ln Z_\psi \ , 
 \label{zetfld} \\
        \zeta_{\widetilde{\psi}} &=& 
                \mu {\partial \over \partial \mu} \bigg\vert_0 
                \ln Z_{\widetilde{\psi}} \ ,
 \label{zetflt}
\end{eqnarray}
and
\begin{equation}
        \zeta_a = \mu {\partial \over \partial \mu}
                \bigg\vert_0 \ln {a \over a_0}
 \label{zetpar}
\end{equation}
(the index `0' indicates that the renormalized fields and parameters are to 
be expressed in terms of their bare counterparts prior to performing the
derivatives with respect to the momentum scale $\mu$).
Note that $\zeta_{\widetilde{\scj}} = \zeta_{\scj} \equiv 0$,
$\zeta_g \equiv d/2-3$ 
as a consequence of Eqs.~(\ref{zmtcho}) and (\ref{galZ}).

The Gell-Mann--Low equation (\ref{calsym}) is readily solved with the method 
of characteristics $\mu \to \mu \ell$; this defines running couplings
as the solutions to the first-order differential RG flow equations
\begin{equation}
        \ell \, {d a (\ell) \over d\ell} = 
        \zeta_a (\ell) \, a(\ell) \, , \
        a(1) = a \ .      
 \label{rgflow}
\end{equation}
The solutions of the partial differential equations (\ref{calsym}) then read
\begin{eqnarray}
        &&\Gamma_{\widetilde{\scj}^k 
        \widetilde{\psi}^r \scj^l \psi^s} \left( \mu, \{ 
        {\bf q},\omega \};\{ a \} \right) =
 \label{solcsy} \\
        &&\quad = \exp \left\{ {1 \over 2} \int_1^\ell 
        \Bigl[ r \, \zeta_{\widetilde{\psi}}(\ell') 
        + s \, \zeta_\psi(\ell') \Bigr] {d \ell' \over \ell'} 
                \right\} \nonumber \\ 
        &&\quad \times \Gamma_{\widetilde{\scj}^k 
        \widetilde{\psi}^r \scj^l \psi^s} 
        \left( \mu \ell, \{ {\bf q},\omega \};
                \{ a(\ell) \} \right) \ . \nonumber
\end{eqnarray}
In the same manner, one can solve the RG equations for the dynamic 
susceptibilities, with the results
\begin{eqnarray}
        &&X \left( \mu, \{ {\bf q},\omega \};
                \{ a \} \right) = 
 \label{solcsx} \\
        &&\quad = X \left( \mu \ell, \{ {\bf q},\omega \};
                \{ a(\ell) \} \right) \ , \nonumber
\end{eqnarray}
and
\begin{eqnarray}
        &&\chi \left( \mu, \{ {\bf q},\omega \};
                \{ a \} \right) = 
 \label{solcsc} \\
        &&\qquad = \exp \left\{ - \int_1^\ell \zeta(\ell') 
        {d \ell' \over \ell'} \right\} \nonumber \\
        &&\qquad \quad \times \chi \left( \mu \ell, 
        \{ {\bf q},\omega \};\{ a(\ell) \} 
                \right) \ , \nonumber
\end{eqnarray}
where, in analogy with Eq.~(\ref{zetfld}),
\begin{equation}
        \zeta= \mu {\partial \over \partial \mu} 
                \bigg\vert_0 \ln Z \ .
 \label{zetflc}
\end{equation}

In terms of the renormalized couplings $\widetilde{u}$, $\widetilde{f}$
and $T$, as defined by Eqs.~(\ref{tildeu})--(\ref{eqT}), we find, using 
the results of subsection \ref{verresp} for the $Z$ factors, the
following explicit results for the zeta functions to one-loop order:
\begin{eqnarray}
        &&\zeta_{\psi}= \frac{4}{(d-2)d(d+2)}\,\widetilde{f}\,\frac{1-T}{T}\,,
 \label{zestpt} \\
        &&\zeta_{\widetilde{\psi}} = -\frac{4}{(d-2)d(d+2)}\,
        \widetilde{f}\,\frac{1-T}{T}\,,
 \label{zesptr} \\
        &&\zeta = 0 + O(\widetilde{u}^2,\widetilde{f}^2)\,,
 \label{zepatr} \\
        &&\zeta_{\lambda}= -\frac{2(d-1)}{(d-2)d}\,\widetilde{f}\,,
 \label{zedtpt} \\
        &&\zeta_{\widetilde{\lambda}} = -\frac{2(d-1)}{(d-2)d}\,
        \frac{\widetilde{f}}{T}+ \frac{4}{(d-2)d(d+2)}\,
                                        \widetilde{f}\,\frac{1-T}{T}\,, \quad  
 \label{zelspt} \\
        &&\zeta_D= -\frac{\widetilde{f}}{4(d+2)}\,,
 \label{zetdpt} \\
        &&\zeta_{\widetilde{D}}= -\frac{1}{4(d+2)}\,\widetilde{f}\,T\,,
 \label{zeltpt} \\                
        &&\zeta_\tau= -2+\frac{\widetilde{u}}{d-2}\,,
 \label{zetrpt} \\      
        &&\zeta_u = -\epsilon+\frac{3}{2}\,\widetilde{u}-\frac{4}{(d-2)d(d+2)}
	\, \widetilde{f}\,\frac{1-T}{T}\,,
 \label{zetupt}
\end{eqnarray}
with $\zeta_g\equiv d/2-3$, identically.
These results enable us now to study the scaling behavior of the 
non-equilibrium model H with dynamical noise in the vicinity of the different
RG fixed points, which are given by the zeros of the appropriate RG beta 
functions ($\{ v \} = \widetilde{u}$, $\widetilde{f}$ and $T$)
\begin{equation}
        \beta_v = \mu {\partial \over \partial \mu} \bigg\vert_0 v \ .
 \label{defbet}
\end{equation}
According to
\begin{equation}
        \ell \, {d v(\ell) \over d \ell} = \beta_v(\{ v(\ell) \}) \ ,
 \label{cpflow}
\end{equation}
these govern the flow of the effective couplings $\widetilde{u}$, 
$\widetilde{f}$, and
$T$ under scale transformations 
$\mu \to \mu \ell$, and the fixed points $\{ v^* \}$ where all 
$\beta_v(\{ v^* \}) = 0$ thus describe scale-invariant regimes.

The RG analysis of the theory then requires the study of the
behavior of three independent coupling constants under the RG flow, 
the static coupling $\widetilde{u}$ defined in (\ref{tildeu}), the 
mode-coupling vertex $\widetilde{f}$ defined in (\ref{tildef}), and 
the parameter $T$ defined in (\ref{eqT}), which denotes
the ratio of the noise temperature of the order
parameter to the noise temperature of the transverse mass current,
with the corresponding beta functions being $\beta_{\widetilde{u}}$,
$\beta_{\widetilde{f}}$ and $\beta_T$. 

Evaluating the Gell-Mann--Low flow equations near an IR-stable RG fixed
point, we may derive the following scaling laws for the two-point correlation
functions of the order parameter and conserved currents, respectively,
\begin{eqnarray}
  C_\psi(\tau,{\bf q},\omega) &=& q^{-2-{\tilde z}_\psi+\eta_\psi} \, 
  {\hat C_\psi}\left( \frac{\tau}{q^{1/\nu}},\frac{\omega}{q^{z_\psi}}\right) 
  \ , \label{opcorr} \\
  C_{\scj}(\tau,{\bf q},\omega) &=& q^{-2-{\tilde z}_{\scj}} \, 
  {\hat C_{\scj}}
\left( \frac{\tau}{q^{1/\nu}},\frac{\omega}{q^{z_{\scj}}}\right)
  \ , \label{jcorr} 
\end{eqnarray}
and for the order parameter susceptibility 
\begin{equation}
  \chi(\tau,{\bf q},\omega) = q^{-2+\eta} \, {\hat \chi}\left( 
  \frac{\tau}{q^{1/\nu}},\frac{\omega}{q^{z_\psi}}\right) \ .
\label{opsusc}
\end{equation}
Here, the different critical exponents are given in terms of the 
following fixed-point values (indicated by a `*') of the zeta functions
\begin{eqnarray}
\label{eta}
\eta&=&-\zeta^*\,,\\
\eta_{\psi}&=&-\zeta^*_{\psi}\,,
\label{etap}\\
\nu^{-1}&=&-\zeta_{\tau}^*\,,
\label{nu}\\
z_{\psi}&=&4+\zeta_{\lambda}^*\,,
\label{zpsi}\\
{\widetilde z}_{\psi}&=&4+\zeta_{\widetilde{\lambda}}^*\,,
\label{ztildepsi}\\
z_{\scj}&=&2+\zeta_D^*\,,
\label{zj}\\
{\widetilde z}_{\scj}&=&2+\zeta_{\widetilde{D}}^*\, .
\label{ztildej}
\end{eqnarray}
The first three exponents correspond in equilibrium to the 
static critical exponents (with $\eta = \eta_\psi$), while the 
last four yield the dynamical critical exponents (again, in
equilibrium $z_\psi = {\widetilde z}_{\psi}$ and $z_{\scj} = 
{\widetilde z}_{\scj}$).

\subsubsection{RG fixed points and critical exponents}
\label{isorgfix}

The relevant RG beta functions, namely $\beta_{\widetilde{u}}$,
$\beta_{\widetilde{f}}$ and $\beta_T$, are given, to one-loop
order, by the following expressions:
\begin{eqnarray}
\label{betau}
\beta_{\widetilde{u}}&=&-\widetilde{u}\left( 
\zeta_{\lambda}-\zeta_{\widetilde{\lambda}}
-\zeta_u\,\right)\nonumber\\
&=&-\widetilde{u}\left(\epsilon-\frac{3}{2}\,\widetilde{u}
+\frac{2(d-1)}{(d-2)d}\,\frac{1-T}{T}\,\widetilde{f}\right)\,,
\\
\label{betaf}
\beta_{\widetilde{f}}&=&-\widetilde{f}
\left(\epsilon+2\zeta_{\lambda}+\zeta_D
-\zeta_{\widetilde{\lambda}}\right)\nonumber\\
&=&-\widetilde{f}\left(\epsilon-\frac{17d^2+14d-48}{4(d-2)d(d+2)}
\,\widetilde{f}\right.\nonumber\\
&&\qquad\quad\mbox{}\left.+\frac{2(d^2+d-4)}{(d-2)d(d+2)}\,
\frac{\widetilde{f}}{T}\right)\,,
\end{eqnarray}
and
\begin{eqnarray}
\beta_T &=& T \left(\zeta_{\widetilde{\lambda}}+\zeta_{D} 
          - \zeta_{\lambda}-\zeta_{\widetilde{D}}\right) \nonumber\\
          &=&-\left[\,\frac{2(d^2+d-4)}{(d-2)d(d+2)}\right.\nonumber\\
      &&\qquad\mbox{}\left. +\frac{T}{4(d+2)}\,\right]\,(1-T)\,\widetilde{f}\,,
\label{betT}
\end{eqnarray}
where we have used Eqs.~(\ref{zedtpt})--(\ref{zetupt}) for the one-loop
Wilson zeta functions. From Eq.~(\ref{betT}) one sees that there exists 
a fixed point either if $T=1$, which is the ordinary equilibrium fixed 
point at which the noise temperatures become equal, or when the
mode coupling vertex $\widetilde{f}=0$, which can yield
a non-equilibrium fixed point (see below).

As a prelude to the study of the behavior of the model at the 
non-equilibrium fixed point and in order to render the analysis more 
transparent, we start by reviewing the results for the equilibrium 
fixed point. Firstly, we notice that in equation (\ref{betT})
one has $\beta_T>0$ for $T>1$ and $\beta_T<0$ for $T<1$.
This implies that in the IR regime $\ell\rightarrow 0$, $T$ increases if 
$T<1$ and it decreases if $T>1$. The equilibrium fixed point is thus
{\em stable} with respect to perturbations in the noise temperature.
Secondly, at this fixed point one obtains
the following values for the coupling constants
$\widetilde{u}^{*}$ and $\widetilde{f}^{*}$,
\begin{eqnarray}
\widetilde{u}^{*}&=&\frac{2}{3}\,\epsilon+O(\epsilon^2)\,,
\label{eqfixu}
\\
\widetilde{f}^{*}&=&\frac{24}{19}\,\epsilon+O(\epsilon^2)\,,
\label{eqfixf}
\end{eqnarray}
where we have performed an expansion around $d=4$ in expressions 
(\ref{betau}) and (\ref{betaf}) and where we have used the fact that
$\beta_{\widetilde{u}}$ is independent of $\widetilde{f}$ when $T=1$.
When these values are substituted in the expressions for the
zeta functions (\ref{zestpt}) to (\ref{zetupt}), one obtains
for the critical exponents, Eqs.~(\ref{eta}) to (\ref{ztildej}), the results
\begin{eqnarray}
&&\eta=\eta_{\psi}=O(\epsilon^2)\,,
\label{eqeta}
\\
&&\nu^{-1}=2-\frac{1}{3}\,\epsilon+O(\epsilon^2)\,,
\label{eqnu}
\\
&&z_{\psi}={\widetilde z}_{\psi}=4-\frac{18}{19}\,\epsilon+O(\epsilon^2)\,,
\label{eqzpsi}
\\
&&z_{\scj}={\widetilde z}_{\scj}=2-\frac{\epsilon}{19}+O(\epsilon^2)\,.
\label{eqzj}
\end{eqnarray}
The values for the static critical exponents are
(due to the existence of detailed balance) the ones one 
obtains if one performs an RG analysis of the 
static model described by the free energy (\ref{hamilt}).
As for the dynamical exponents, whereas $z_{\scj}$
shows a small negative correction due to the dynamical
coupling between the two modes, the dynamical
exponent $z_{\psi}$, which describes the behavior
of the characteristic frequency of the order
parameter $\omega_{\psi}(q)\propto q^{z_{\psi}}$,
displays a strong suppression already in $d=3$ ($\epsilon=1$).
This can be understood from the fact that in the absence of 
dynamical coupling and at the critical point, the dynamics
of the conserved current is purely diffusive and therefore 
much faster than the dynamics of the also {\em conserved} order 
parameter which shows the characteristic critical slowing down 
\cite{nocritslow}. The existence of dynamical coupling between 
the two modes will then provide the slower order parameter field with 
an additional and faster decay channel, so that $z_\psi < 4$. Notice
furthermore that Eq.~(\ref{betaf}) implies for {\em any} fixed 
point with non-trivial and finite mode-coupling
$0 < {\widetilde f}^* < \infty$, in equilibrium (where 
$\zeta_\lambda = \zeta_{\widetilde{\lambda}}$) the relation
$\zeta_\lambda^* + \zeta_D^* = - \epsilon = d-4$, i.e., the exponent
identity
\begin{equation}
   z_\psi + z_{\scj} \equiv 6 - \epsilon = d + 2 \ ,
\label{eqexid}
\end{equation}
which is of course satisfied by the explicit one-loop results
(\ref{eqzpsi}) and (\ref{eqzj}).

If one now takes $\widetilde{f}=0$ and $T$ finite in Eqs.~(\ref{betau}) 
to (\ref{betT}), one obtains a line of fixed points (i.e., one fixed 
point for each value of $T$), provided that
$\widetilde{u}=\frac{2}{3}\,\epsilon+O(\epsilon^2)$.
This line of fixed points corresponds to a system
where the two modes are completely decoupled, as can
be seen from the computation of the critical exponents.
One has
\begin{eqnarray}
&&\eta=\eta_{\psi}=O(\epsilon^2)\,,
\label{eqeta2}
\\
&&\nu^{-1}=2-\frac{1}{3}\,\epsilon+O(\epsilon^2)\,,
\label{eqnu2}
\\
&&z_{\psi}={\widetilde z}_{\psi}=4+O(\epsilon^2)\,,
\label{eqzpsi2}
\\
&&z_{\scj}={\widetilde z}_{\scj}\equiv2\,.
\label{eqzj2}
\end{eqnarray}
with the last equality holding identically, given that 
$\widetilde{f}=0$. The static critical exponents and the
dynamical exponents $z_{\psi}$, ${\widetilde z}_{\psi}$ are simply
the exponents one would obtain for the equilibrium purely relaxational
critical dynamics model B with conserved order parameter, with the 
$O(\epsilon^2)$ term representing two-loop contributions coming from 
the static vertex. For the dynamic critical exponent, one knows that
in fact
\begin{equation}
  z_\psi \equiv 4 - \eta
\label{modbz}
\end{equation}
exactly \cite{hohhal,dynfun}. However, from Eq.~(\ref{betaf}) we see
that the model B fixed point is IR-{\em unstable} against perturbations 
in $\widetilde{f}$ since $\beta_{\widetilde{f}}<0$ for small $\widetilde{f}$.

The cases $\widetilde{f}=0$ with $T=0$ or $T=\infty$ require
extra care, as one needs to consider the flow of 
$\bar{f}'=\widetilde{f}/T$ in the first case, and 
the flow of $\bar{f}=\widetilde{f}\,T$ in the second case,
since these two quantities might, respectively, have a finite
value at the hypothetical fixed points.
If $\widetilde{f}=0$ and $T=0$, then $\beta_{\bar{f}'}=-\epsilon \bar{f}'$
with $\bar{f}'=\widetilde{f}/T$,
which shows that a fixed point cannot exist for any finite
value of $\bar{f}'$. Physically, the absence of a fixed point
is again due to the fact that the order parameter displays critical 
slowing down compared to its diffusive relaxation
away from the critical point.  The conserved current is,
on the other hand, always governed by faster, diffusive 
dynamics. If the conserved current were to be placed at infinite temperature, 
it would be slaving the order parameter, which is impossible.
Thus no fixed point corresponding to such a situation exists, in contrast 
with the isotropic non-equilibrium SSS model \cite{uwezol}. 
There, the order parameter is not conserved, thus permitting a $T=0$ 
fixed point (albeit an unstable one) where the purely diffusive (as there
is no feedback from the order parameter) conserved variable actually 
becomes the slower dynamical mode, with $z_\psi = 2 -\epsilon/2+O(\epsilon^2)$
for the order parameter \cite{uwezol,uwejai}.

\begin{figure}
  \centerline{\epsfxsize 0.75\columnwidth \epsfbox{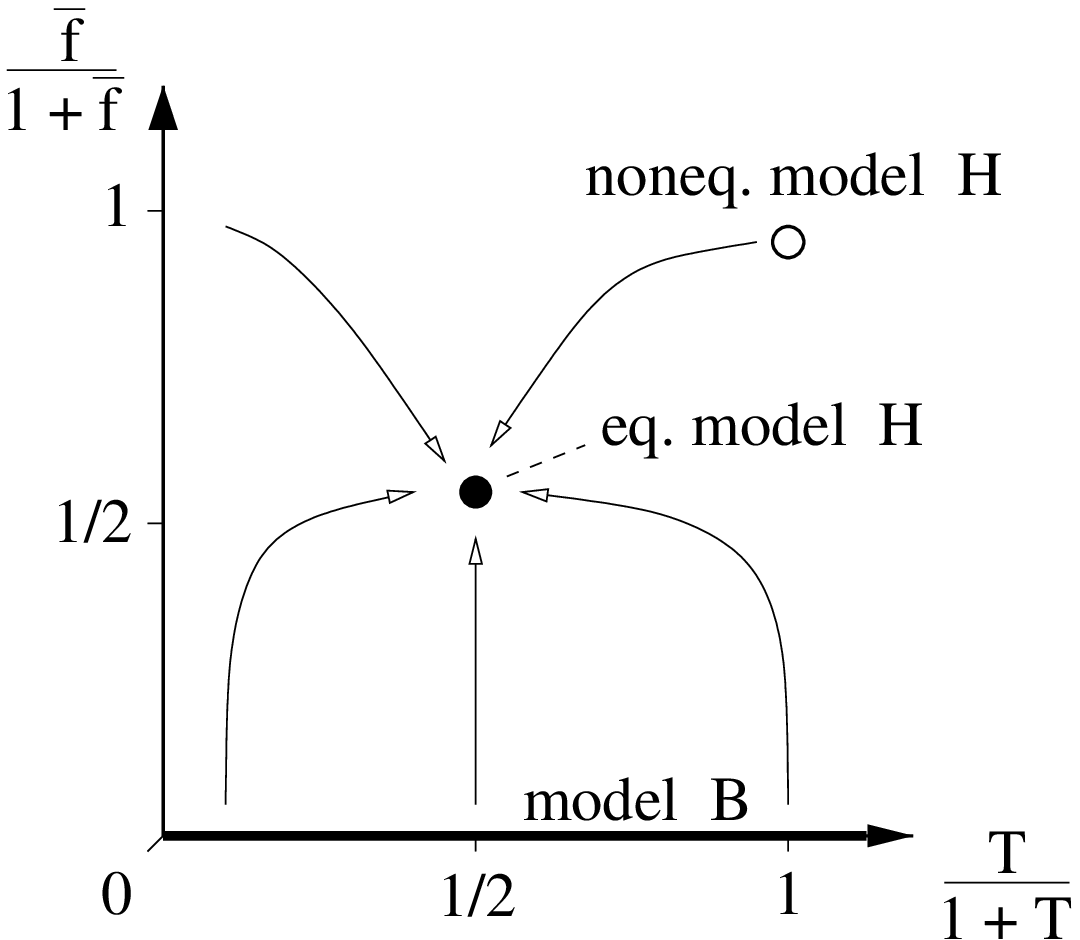}}
\caption{Equilibrium and non-equilibrium fixed points for the isotropic
  non-equilibrium model H, plotted for $d = 3$ ($\epsilon = 1$).} The
  equilibrium fixed point in the center of the flow diagram is 
  infrared-stable. The model B fixed line is unstable with respect to
  the mode-coupling, the non-equilibrium fixed point at $T = \infty$ is
  unstable in the $T$ direction.
\label{hfig}
\end{figure}

If $\widetilde{f}=0$ and $T=\infty$, one finds, on the other hand, for the
RG beta function corresponding to the new coupling
$\bar{f}=\widetilde{f} \, T$:
\begin{equation}
\beta_{\bar{f}}=-\left(\epsilon-\frac{1}{4(d+2)}\,\bar{f}\right)\,
\bar{f}\,,
\label{betabarf}
\end{equation}
which has a {\em non-equilibrium} 
fixed point for $\bar{f}^*=4(d+2)\epsilon$. It is
also easy to see that, since $\beta_{\bar{f}}<0$ for $\bar{f}<\bar{f}^*$,
$\beta_{\bar{f}}>0$ for $\bar{f}>\bar{f}^*$, this genuine non-equilibrium 
fixed point is IR-{\em stable} with respect to perturbations on the value 
of $\bar{f}$. However, it is {\em unstable} against perturbations on the 
value of $T$, since $\beta_T>0$ for large $T$. The values of the critical 
exponents at this unstable fixed point are given by
\begin{eqnarray}
&&\eta=\eta_{\psi}=O(\epsilon^2)\,,
\label{eqeta3}
\\
&&\nu^{-1}=2-\frac{1}{3}\,\epsilon+O(\epsilon^2)\,,
\label{eqnu3}
\\
&&z_{\psi}={\widetilde z}_{\psi}=4+O(\epsilon^2)\,,
\label{eqzpsi3}
\\
&&z_{\scj}=2\,,
\label{eqzj3}
\\
&&{\widetilde z}_{\scj}=2-\epsilon+O(\epsilon^2)=d-2+O(\epsilon^2)\, .
\label{eqzjtilde3}
\end{eqnarray}
Eq.~(\ref{eqzj3}) holds to all orders given the structure of the vertices 
at this fixed point. The $O(\epsilon^2)$ corrections to the static critical 
exponents and to $z_{\psi}={\widetilde z}_{\psi}$ stem from the two-loop 
contributions induced by the {\em static} vertex only. We thus expect the
exact relation (\ref{modbz}) to hold here as well. Moreover, 
as we have $\zeta_\lambda=\zeta_{\widetilde{\lambda}}$ at
$T=\infty$, the very
existence of {\em any} non-trivial, finite fixed point $0 < \bar{f} < \infty$
demands that $\zeta_\lambda^* + \zeta_{\widetilde D}^* = - \epsilon$, i.e.,
$z_\psi + {\widetilde z}_{\scj} = 6 -\epsilon = d+2$, which generalizes the 
corresponding equilibrium scaling relation (\ref{eqexid}). With 
$\zeta_\lambda = - \eta_{\psi}$ we therefore arrive at
\begin{equation}
  {\widetilde z}_{\scj} = d - 2 + \eta_\psi \ .
\label{annsex}
\end{equation}
One sees that this fixed point is characterized by model
B exponents for the order parameter and by anomalous
noise correlations for the conserved current, giving
rise to a value of $\widetilde{z}_{\scj}\neq2$ for $d < d_c = 4$.
Again, the model B exponents obtained for the order parameter 
can be understood from the fact that the conserved current is 
effectively at zero temperature, and therefore its dynamics does not 
influence the dynamics of the order parameter, being however affected 
by it through the residual one-way coupling between these two dynamic 
variables. Such a fixed point is also present in the non-equilibrium 
SSS model \cite{uwezol}, with similar model A-type behavior for the 
order parameter, and anomalous noise properties for the dynamically
coupled conserved fields. In the present context its existence is due 
to the fact that here the slower variable is at higher temperature, 
slaving the faster conserved modes, and suffering no feedback from 
the latter. The characteristics of this novel, unstable non-equilibrium 
fixed point should be contrasted with the pure model B fixed point 
discussed above where there is no coupling at all: $\widetilde{f}=0$,
and where the temperature ratio $T$, albeit finite, does not matter, 
for the two Langevin equations are fully decoupled. The RG fixed point
structure, and their stability is summarized in Fig.~\ref{hfig}.

\section{The anisotropic non-equilibrium model H}
\label{anisoH}

In this section, we study the critical behavior of our non-equilibrium 
version for model H with dynamical noise, as defined through 
Eqs.~(\ref{hoplan}) and (\ref{samlan}) and by the {\em anisotropic} 
noise correlators (\ref{animodelH1}) and (\ref{animodelH2}).
We start by computing the $T_c$ shift from the static susceptibility.
As a consequence of the spatially anisotropic conserved noise with 
$T_0^\perp < T_0^\parallel$, it turns out that the transverse momentum 
space sector with {\em lower} noise temperature softens first.
Thus, at the critical point, the longitudinal sector remains uncritical 
(`stiff'), similar to equilibrium anisotropic elastic phase transitions 
\cite{elphtr} and at Lifshitz points \cite{lifsht}.
In Sec.~\ref{mdjren}, we turn to the perturbational renormalization of
the two-temperature non-equilibrium model J to one-loop order, and finally
discuss the resulting RG flow equations.

\subsection{Dynamic field theory and the anisotropic ${\bf T_c}$ shift}
 \label{atcshf}

The dynamic field theory which describes the anisotropic
model H has already been presented in Sec.~\ref{modeqs},
the dynamic functional $J$ being given by Eqs.~(\ref{harmfun}) to 
(\ref{mctfun}), with the choice (\ref{animodelH1}) and (\ref{animodelH2}) 
for the noise correlators. However, the computation of the shift in the 
critical temperature shows that, as it stands, this theory is not fully 
consistent. The shift in the critical temperature is determined,
as in the isotropic theory, by the condition 
$\chi_0^{-1}(\vec{q},\omega)=0$ in the limit 
$\vec{q}\rightarrow\vec{0}$, $\omega\rightarrow 0$,
where $\chi_0(\vec{q},\omega)$ is the order parameter
dynamic susceptibility, which is given in terms 
of the relevant vertex functions by Eq.~(\ref{opvsus}).

From the ensuing expression (to one-loop order), we may determine the 
fluctuation-induced shift of the critical temperature.
Because of the dynamic anisotropy appearing in the noise correlators 
(\ref{animodelH1}) and (\ref{animodelH2}), however, 
the result depends on how the limit 
${\bf q} \to {\bf 0}$ is taken; upon defining 
$q_\parallel = q \cos \Theta$ and
$q_\perp = q \sin \Theta$ and with
$T_0^{\parallel}=(\widetilde{\lambda}_0^{
\parallel}/\lambda_0)\,(D_0/\widetilde{D}_{0}^{\parallel})$,
$T_0^{\perp}=(\widetilde{\lambda}_0^{
\perp}/\lambda_0)\,(D_0/\widetilde{D}_{0}^{\perp})$
denoting the temperature ratios in the different sectors, we find
\begin{eqnarray}
\label{anicrit}
        &&r_{0c}(\Theta) =\frac{d_{\parallel}}{d}
       \frac{\widetilde{D}_{0}^{\parallel}}{D_0}\,\left(-{1\over 2} 
         \, u_0 T_0^\parallel \int_k {1 \over r_{0c} + k^2}\right.\\
        &&\mbox{}
        \left.+{g_0^2(d+1) \over  \lambda_0^2 (d+2)}(T_0^{\parallel}-1)\,
         \int_k {1 \over k^2(r_{0c} + D_0/\lambda_0 + k^2)}
                \right)\nonumber\\
&&\mbox{}+\frac{d_{\perp}}{d}
       \frac{\widetilde{D}_{0}^{\perp}}{D_0}\,\left(-{1\over 2} 
         \, u_0 T_0^\perp \int_k {1 \over r_{0c} + k^2}\right.\nonumber\\
        &&\mbox{}
        \left.+{g_0^2(d+1) \over \lambda_0^2 (d+2)}(T_0^{\perp}-1)\,
         \int_k {1 \over k^2(r_{0c} + D_0/\lambda_0 + k^2)}
                \right)\nonumber\\
&&\mbox{}-{2g_0^2 \over \lambda_0^2 d(d+2)}\,
         \int_k {1 \over k^2(r_{0c} + D_0/\lambda_0 + k^2)}
\nonumber\\
        &&\mbox{}\times\left[\left(
\frac{\widetilde{\lambda}_0^\parallel}{\lambda_0}
-\frac{\widetilde{D}_0^\parallel}{D_0}\right)\,\cos^2\Theta
+\left(\frac{\widetilde{\lambda}_0^\perp}{\lambda_0}
-\frac{\widetilde{D}_0^\perp}{D_0}\right)\,\sin^2\Theta\,\right] \ ,
\nonumber 
\end{eqnarray}
in contrast with Eq.~(\ref{r0c}) for the isotropic model H.
Here, $d_\parallel$ and $d_\perp$ are, respectively, the dimensions
of the parallel and transverse subspaces, with $d=d_\parallel+d_\perp$.
As $T_c = T_c^0 + r_{0c}$, the phase transition will occur at the maximum of 
the function $r_{0c}(\Theta)$, which for 
$\left(\frac{\widetilde{\lambda}_0^\perp}{\lambda_0}
-\frac{\widetilde{D}_0^\perp}{D_0}\right)<\left(
\frac{\widetilde{\lambda}_0^\parallel}{\lambda_0}
-\frac{\widetilde{D}_0^\parallel}{D_0}\right)$, or, equivalently, 
$T_0^\perp - 1 < (T_0^\parallel - 1) 
\frac{\widetilde{D}_0^\parallel}{\widetilde{D}_0^\perp}$, occurs at 
$\Theta = \pi / 2$. The $d_\perp$-dimensional transverse sector in 
momentum space thus softens first, and the true $T_c$ shift is given by
\begin{eqnarray}
\label{anicrit2}
        &&r_{0c}\left(\frac{\pi}{2}\right) =\frac{d_{\parallel}}{d}
       \frac{\widetilde{D}_{0}^{\parallel}}{D_0}\,\left(-{1\over 2} 
         \, u_0 T_0^\parallel \int_k {1 \over r_{0c} + k^2}\right.\\
        &&\mbox{}
        \left.+{g_0^2(d+1) \over  \lambda_0^2 (d+2)}(T_0^{\parallel}-1)\,
         \int_k {1 \over k^2(r_{0c} + D_0/\lambda_0 + k^2)}
                \right)\nonumber\\
&&\mbox{}+\frac{d_{\perp}}{d}
       \frac{\widetilde{D}_{0}^{\perp}}{D_0}\,\left(-{1\over 2} 
         \, u_0 T_0^\perp \!\int_k {1 \over r_{0c} + k^2}
         +{g_0^2\over \lambda_0^2 (d+2)}(T_0^{\perp}-1)\,
        \right.\nonumber\\
        &&\mbox{}\left.\times\left(d+1-\frac{2}{d_\perp}\right)
         \int_k {1 \over k^2(r_{0c} + D_0/\lambda_0 + k^2)}
                \right)\,,\nonumber
\label{tctint}
\end{eqnarray}
where again this non-universal quantity must be computed with a finite
UV cutoff. For $T_0^\parallel = T_0^\perp = T_0$, we recover the result 
(\ref{r0c}) for the isotropic model. Notice that dynamical anisotropy 
($T_0^\parallel \not= T_0^\perp$), {\em combined} with the reversible 
mode-coupling terms, has a very drastic effect here: It renders the system 
soft only in the momentum subspace with {\em lower} effective noise 
temperature. This effect has a simple physical interpretation: The $T_c$ 
shift is due to thermal fluctuations, which are reduced in the transverse 
sector ($T_0^\perp < T_0^\parallel$), and therefore lead to a comparatively 
stronger downwards shift in the longitudinal sector. This result is 
completely analogous to our earlier findings for the non-equilibrium model 
J (describing isotropic ferromagnets) with anisotropic noise \cite{uwejai}.

In order to characterize the critical properties of our model, we may neglect
terms $\propto q_\parallel^4$ in the stiff momentum space sector, because
$\tau_0^\parallel = r_0 - r_{0c}(\Theta = 0)$ remains positive at the phase 
transition where $\tau_0^\perp = r_0 - r_{0c}(\Theta = \pi / 2)$ vanishes.
In analogy with the situation at anisotropic elastic structural phase 
transitions \cite{elphtr}, or with Lifshitz points in magnetic systems with
competing interactions \cite{lifsht}, as well as driven diffusive systems
\cite{bearev}, we thus have to scale the soft and stiff wave vector components 
differently, $[q_\perp] = \mu$, whereas $[q_\parallel] = [q_\perp]^2 = \mu^2$.
Consequently, while $[{\widetilde \lambda}_0^\perp] = \mu^0$ and 
$[{\widetilde D}_0^\perp] = \mu^2$, if we choose $[\omega] = \mu^4$, we find 
for the longitudinal noise strengths the scaling dimensions 
$[{\widetilde \lambda}_0^\parallel]=\mu^{-2}$ and 
$[{\widetilde D}_0^\parallel]=\mu^{0}$, which implies that they become 
{\em irrelevant} under scale transformations. Allowing for distinct 
couplings in the different sectors, one finds in the same manner that the 
ratios $[\lambda_0^\parallel / \lambda_0^\perp] = [D_0^\parallel / D_0^\perp]
= [\lambda_0^\parallel u_0^\parallel / \lambda_0^\perp u_0^\perp] = \mu^{-2}$
and $[g_0^\parallel / g_0^\perp] = \mu^{-1}$ all have negative scaling 
dimension. Thus, for an investigation of the asymptotic critical behavior, 
the longitudinal parameters may be neglected as compared to their transverse 
counterparts, and can all be set to zero in the {\em effective} dynamic 
functional.

Upon rescaling the fields according to $\psi_0\!\to\! 
({\widetilde \lambda}_0^\perp/\lambda_0^\perp)^{1/2}$ $\psi_0$, 
${\widetilde \psi}_0 \to 
(\lambda_0^\perp/{\widetilde \lambda}_0^\perp)^{1/2} {\widetilde \psi}_0$,
$\scj^{\alpha}_0 \to 
({\widetilde D}_0^\perp/D_0^\perp)^{1/2} \scj^{\alpha}_0$
and  $\widetilde{\scj}_0^{\alpha} \to 
(D_0^\perp/{\widetilde D}_0^\perp)^{1/2} 
\widetilde{\scj}_0^\alpha$ and defining
\begin{eqnarray}
        &&c_0 = {\lambda_0^\parallel \over \lambda_0^\perp} \, \tau_0^\parallel
        \; , \quad 
        {\widetilde u}_0 = {{\widetilde \lambda}_0^\perp \over \lambda_0^\perp}
        \, u_0^\perp \; , \quad \nonumber\\
        &&g_0 = \sqrt{{\widetilde D}_0^\perp \over 
                D_0^\perp} \, g_0^\perp \ ,\quad
        \widetilde{g}_0={{\widetilde \lambda}_0^\perp \over 
           \lambda_0^\perp}\sqrt{D_0^\perp \over 
                \widetilde{D}_0^\perp} \, g_0^\perp \ ,
 \label{nejcpl}
\end{eqnarray}
and omitting the labels `$\perp$' again for $\lambda_0$ and $r_0$, the 
ensuing {\em effective} Langevin equations of motion become
\begin{eqnarray}
        &&{\partial \psi_0 \over \partial t} =
        \lambda_0 \left[ c_0 \nabla_\parallel^2 + 
        \nabla_\perp^2 (r_0 - \nabla_\perp^2) \right] \psi_0 +
 \label{nejlan} \\
        &&\ \qquad + \lambda_0 {{\widetilde u}_0 \over 6} \nabla_\perp^2 
                \psi_0^3-g_0 
       \nabla_\perp\psi_0\cdot\vecj_0+\eta \ , \nonumber
\label{psilani}
\end{eqnarray}
and
\begin{eqnarray}
        {\partial \mbox{\bf \j}_0\over \partial t}
        \!\!&=&\!\! {\cal T}\left[D_0\nabla^2_{\perp}\vecj_0
        +\widetilde{g}_0\nabla_\perp\psi_0(r_0-\nabla^2_\perp)\psi_0
        +\vec{\zeta}\right],
\label{samlani}
\end{eqnarray}
where the transverse projector in the soft ($\perp$) subspace ${\cal T}$ is 
given, in Fourier space, by the expression ${\cal T}^{\alpha\beta}(\vec{k})=
\delta^{\alpha\beta}-k^\alpha_\perp k^\beta_\perp/k^2_\perp$.
The noise correlators in turn read
\begin{equation}
        \langle \eta(\vec{k},\nu) \eta(\vec{k}',\nu') 
        \rangle = 2 {\widetilde \lambda}_0 \,k_\perp^2\, 
        \delta(\vec{k} + \vec{k}') \,
                        \delta( \nu + \nu') 
 \label{sopnoani}
\end{equation}
and
\begin{eqnarray}
        \langle \zeta^{\alpha}(\vec{k},\nu) 
                \zeta^{\beta}(\vec{k}',\nu') \rangle &=&
        2 {\widetilde D}_0\,k_\perp^2\,
                \delta(\vec{k} +\vec{k}') \,
                \delta( \nu + \nu') \nonumber \\
        &&\mbox{}\times \left(\delta^{\alpha\beta}
        -\frac{k_\perp^\alpha k^\beta_\perp}{k_\perp^2}\right) ,
 \label{samnoani}
\end{eqnarray}
where again, for convenience, we have used the Fourier space representation.
These equations define the {\em two-tem\-pe\-rature non-equili\-bri\-um 
model H}. In order to perform the RG analysis, one represents these equations
in the form of a dynamic functional, precisely as in Sec.~\ref{modeqs} above.

We emphasize the fact that the anisotropy of the $T_c$ shift in 
Eq.~(\ref{anicrit}) only occurs in the contribution $\propto g_0^2$,
i.e., the anisotropy in the $T_c$ shift is due to the purely dynamical 
mode-coupling terms. In the non-equilibrium model B with 
dynamical anisotropy \cite{bearoy}, the criticality condition for the 
response function remains isotropic, at least to one-loop order.
Thus, if one does not assume different critical temperatures in the purely 
diffusive non-linear Langevin equation to begin with, these are {\em not} 
generated, and one is not immediately led to the two-temperature model B 
as the correct effective theory for the phase transition. In the presence
of reversible mode-couplings, however, anisotropic noise correlations,
specifically for a conserved order parameter field, have a much more 
drastic effect: For both models J \cite{uwejai} and model H such violations
of the detailed-balance conditions render the system inherently anisotropic
at criticality, and certainly prevent any restoration of the equilibrium
critical properties.

\subsection{Renormalization of the two-temperature model H}
\label{mdjren}

We start by noticing that, as with the isotropic non-equi\-li\-brium model
H, the two-temperature model H, being a genuinely non-equi\-li\-brium model
as well, does not allow us to invoke a fluc\-tua\-tion-dissipation theorem 
in order to relate vertex and response function renormalizations, and we 
have to compute almost all the $Z$ factors independently. These consist of 
the wave function renormalization factors $Z_{\psi}$, $Z_{\widetilde{\psi}}$,
$Z_{\scj}$, $Z_{\widetilde{\scj}}$, as given by (\ref{conren}) and 
(\ref{fldren}), and the coupling constant renormalizations, which we define 
here through
\begin{eqnarray}
\lambda&=&Z_{\lambda}\,\lambda_0 \,,
\label{RGlambdaani}\\
c&=&Z_{c}\,c_0 \,, 
\label{RGcani}\\
D&=&Z_{D}\,D_0\,\mu^{-2} \,,
\label{RGDani}\\
\tau&=&Z_{\tau}\,\tau_0\,\mu^{-2}\;\;\mbox{with}\;\;\tau_0=r_0-r_{0c} \,,
\label{RGtauani}\\
{\widetilde u}&=&Z_{\widetilde u}\,\widetilde{u}_0\,
A(d_\parallel,d_\perp)\,\mu^{d+d_\parallel-4} \,,
\label{RGuani}\\
g&=&Z_g^{1/2}\,g_0\,A(d_\parallel,d_\perp)^{1/2}\,\mu^{(d+d_\parallel)/2-3} \,,
\label{RGgani}\\
\widetilde{g}&=&Z_{\widetilde{g}}^{1/2}\,\widetilde{g}_0\,
A(d_\parallel,d_\perp)^{1/2}\,\mu^{(d+d_\parallel)/2-3}\,,
\label{RGgani2}
\end{eqnarray}
where $A(d_\parallel,d_\perp)$ is given by
\begin{equation}
        A(d_\parallel,d_\perp) = {\Gamma(3-d/2-d_\parallel/2) \, \Gamma(d/2) 
        \over 2^{d-1} \pi^{d/2} \, \Gamma(d_\perp/2)} \ .
 \label{adpatr}
\end{equation}
Again, this factor is included because it appears in subsequent formulas.

As before, Eq.~(\ref{amverf}) implies that the relations (\ref{zmtrel1}) 
and (\ref{zmtrel2}) for the wave function renormalizations hold. 
Furthermore, the vertex structure of the model leads to
\begin{equation}
 \Gamma_{0 \,\widetilde{\psi} \psi}\,
(\vec{q}_\parallel,\vec{q}_{\perp} = {\bf 0},\omega=0) =\lambda_0
\,c_0\,q_\parallel^2\,,
 \label{anilc}
\end{equation}
which must hold to all orders. This entails that
\begin{equation}
Z_{\lambda}\,Z_c\equiv 1,
\label{Zlc}
\end{equation}
which leaves us with one $Z$ factor less to determine.
One has thus a total of seven independent renormalization
factors. As the vertex functions or their derivatives
which we must render finite are the same as for the 
isotropic model H, given at the beginning of Sec.~\ref{verresp}, 
with the exception that one has to substitute 
the derivatives $\partial_{q}$ with respect to $\vec{q}$
by derivatives $\partial_{q_\perp}$ with respect to $\vec{q}_{\perp}$,
one sees that the renormalization conditions 
on these vertex functions determine all the renormalization factors, 
i.e., there is no freedom left, as opposed to the isotropic model
H, to arbitrarily fix one of the renormalizations.
To these renormalization factors, one adds, as above, the 
renormalization factor $Z$, defined in Eq.~(\ref{rensus}), necessary to 
render the order parameter susceptibility finite. It is determined 
by the condition that $\partial_{q_\perp^2} \chi^{-1}({\bf q},\omega)
\vert_{\rm NP}$ be UV-finite \cite{susani}. 
Furthermore, Galilean invariance still holds in the two-temperature
model (see Appendix \ref{appA}), which entails that 
\begin{equation}
Z_{g}\,Z_{\scj}\equiv 1,
\label{ZGalani}
\end{equation}
to all orders, although, as pointed above, one can now no longer
suppose that $Z_{\scj}= 1$.

Next we introduce the coupling constants
\begin{eqnarray}
v_0&=&\frac{\widetilde{u}_0}{c_0^{d_\parallel/2}} \ ,
\label{tildev0ani}
\\
f_0&=&\frac{g_0^2}{\lambda_0 D_0 c_0^{d_\parallel/2}} \ ,
\label{f0ani}
\\
\bar{f}_0&=&\frac{g_0\widetilde{g}_0}{\lambda_0 D_0 c_0^{d_\parallel/2}} \ ,
\label{barf0ani} 
\\
\widetilde{f}_0&=&\frac{\widetilde{g}_0^2}{\lambda_0 D_0 c_0^{d_\parallel/2}}
\ ,
\label{tildef0ani}
\end{eqnarray}
and their renormalized counterparts
\begin{eqnarray}
v&=&Z_{\widetilde{u}}\,Z_\lambda^{d_\parallel/2}\,v_0\,A(d_\parallel,d_\perp)\,
\mu^{-\epsilon} \ ,
\label{tildevani}
\\
f&=&\frac{Z_g}{Z_{\lambda}^{1-d_\parallel/2}Z_D}\,f_0\,A(d_\parallel,d_\perp)\,
\mu^{-\epsilon} \ ,
\label{fani}
\\
\bar{f}&=&\frac{(Z_gZ_{\widetilde{g}})^{1/2}}{
Z_{\lambda}^{1-d_\parallel/2}Z_D}\,\bar{f}_0\,A(d_\parallel,d_\perp)\,
\mu^{-\epsilon} \ ,
\label{barfani}
\\
\widetilde{f}&=&
\frac{Z_{\widetilde{g}}}{Z_{\lambda}^{1-d_\parallel/2}Z_D}
\,\widetilde{f}_0\,A(d_\parallel,d_\perp)\,
\mu^{-\epsilon}\ ,
\label{tildefani}
\end{eqnarray}
where $\epsilon=4-d-d_\parallel$. Notice that all these renormalized
non-linear couplings become scale-invariant at the {\em reduced} upper
critical dimension
\begin{equation}
  d_c(d_\parallel) = 4 - d_\parallel \ .
\end{equation}
Such a lowering of the critical dimension is typical of models with
anisotropic scaling \cite{elphtr,lifsht,bearev}, for the fluctuations 
are critical merely in the transverse sector.
Employing the renormalization conditions, one then obtains 
to one-loop order the following results for the $Z$ factors
\begin{eqnarray}
Z_{\lambda}&=&1+
\frac{6(d_\perp-2)(d_\perp-1)}{(d-2)d_\perp(d_\perp+2)}\,\bar{f}_0\,
\frac{A(d_\parallel,d_\perp)}{\epsilon}\,\mu^{-\epsilon}
\,,
\label{Zlani}
\\
Z_{\tau}&=&1
-\frac{2}{d+d_\parallel-2}\,
\left(\,\frac{v_0}{2}-\frac{d_\perp-1}{d_\perp}\,\bar{f}_0\,\right)
\frac{A(d_\parallel,d_\perp)}{\epsilon}\,\mu^{-\epsilon}\nonumber\\
&&\mbox{}-\frac{6(d_\perp-2)(d_\perp-1)}{(d-2)d_\perp(d_\perp+2)}
\,\bar{f}_0\,\frac{A(d_\parallel,d_\perp)}{\epsilon}\,\mu^{-\epsilon}
\,,
\label{Ztani}
\\
Z_{D}&=&1+\frac{1}{4(d_\perp+2)}\,\bar{f}_0\,
\frac{A(d_\parallel,d_\perp)}{\epsilon}\,\mu^{-\epsilon} \,,
\label{ZtDani}
\\
Z_{\psi}&=&1-
\frac{2(d_\perp-1)}{(d-2)d_\perp}\,f_0\,
\frac{A(d_\parallel,d_\perp)}{\epsilon}\,\mu^{-\epsilon}\\
&&
\mbox{}+\frac{6(d_\perp-2)(d_\perp-1)}{(d-2)d_\perp(d_\perp+2)}\,\bar{f}_0\,
\frac{A(d_\parallel,d_\perp)}{\epsilon}\,\mu^{-\epsilon}\,,\nonumber
\label{Zpsiani}
\\
Z_{\scj}&=&1+\frac{1}{4(d_\perp+2)}\,\bar{f}_0\,
\frac{A(d_\parallel,d_\perp)}{\epsilon}\,\mu^{-\epsilon}\nonumber\\
&&
\mbox{}-\frac{d}{4d_\perp(d_\perp+2)}\tilde{f}_0\,
\frac{A(d_\parallel,d_\perp)}{\epsilon}\,\mu^{-\epsilon}
\,,
\label{Zjani}
\\
Z_{\widetilde{u}}&=&1-\frac{3}{2}\,v_0\,
\frac{A(d_\parallel,d_\perp)}{\epsilon}\,\mu^{-\epsilon}\\
&&+\frac{2(d_\perp-1)}{(d-2)d_\perp}\,f_0\,
\frac{A(d_\parallel,d_\perp)}{\epsilon}\,
\mu^{-\epsilon}\nonumber\\
&&\!\!+\mbox{}\frac{d_\perp-1}{d_\perp}\,
\left(1-\frac{12(d_\perp-2)}{(d-2)(d_\perp+2)}\,\right)\!
\tilde{f}_0\,\frac{A(d_\parallel,d_\perp)}{\epsilon}\,
\mu^{-\epsilon}\nonumber\,,
\label{Zuani}
\\
Z_{\widetilde{g}}&=&1+\frac{4}{(d-2)d_\perp}\!\left(\!d_\perp-1
-\frac{2}{d_\perp+2}\right)\!f_0\,
\frac{A(d_\parallel,d_\perp)}{\epsilon}\,
\mu^{-\epsilon}\nonumber\\
&&\mbox{}-\frac{4}{d_\perp(d_\perp+2)}\,
\left(3(d_\perp-1)\frac{d_\perp-2}{d-2}-\frac{d_\perp+16}{16}\right)
\nonumber\\
&&\mbox{}\times
\bar{f}_0\,\frac{A(d_\parallel,d_\perp)}{\epsilon}\,
\mu^{-\epsilon}\nonumber\\
&&\mbox{}-\frac{d}{4d_\perp(d_\perp+2)}\,\tilde{f}_0\,
\frac{A(d_\parallel,d_\perp)}{\epsilon}\,
\mu^{-\epsilon}\,,
\label{Zgani}
\end{eqnarray}
with $Z_{g}=Z_{\scj}^{-1}$, a result which can be confirmed explicitly 
to one-loop order from the renormalization condition of
$\partial_{q^{a}_{\perp}} 
\Gamma_{ \, \widetilde{\psi}\psi\,\scj^{\alpha}}(\vec{-q-p},
-\omega;\vec{q},\omega;\vec{p},0)\mid_{NP}$.
Subsequently, rendering $\partial_{q_\perp^2} \, 
\chi^{-1}({\bf q},\omega = 0) \vert_{\rm NP}$ 
UV-finite yields the additional $Z$ factor for the response function
\begin{eqnarray}
Z\!&=&\!1+
\frac{4(d_\perp-4)(d_\perp-1)}{(d-2)d_\perp(d_\perp+2)}\,\bar{f}_0\,
\frac{A(d_\parallel,d_\perp)}{\epsilon}\,
\mu^{-\epsilon}\,.
\end{eqnarray}
These $Z$ factors can now be used to compute the relevant
beta functions and Wilson zeta functions of the theory.

\subsection{Discussion of the RG flow equations}
 \label{mdjflo}

Being in possession of the expressions for the $Z$ factors to one-loop
order, one can, in an analogous manner to what was done in sections 
\ref{isorgver} and \ref{isorgfix}, compute the relevant Wilson
zeta functions, also to one-loop order. These are given by
\begin{eqnarray}
        &&\zeta_{\psi}= \frac{2(d_{\perp}-1)}{(d-2)d_{\perp}}
\,f\,-\,6\frac{(d_{\perp}-2)(d_{\perp}-1)}{(d-2)d_{\perp}(d_{\perp}+2)}\,
\bar{f}\,,
 \label{zestptani} \\
        &&\zeta_{\scj} = -\frac{1}{4(d_{\perp}+2)}\,\bar{f}\,+\,
        \frac{d}{4d_{\perp}(d_{\perp}+2)}\,\widetilde{f}\,,
 \label{zesptrani} \\
        &&\zeta = -\frac{4(d_\perp-4)(d_\perp-1)}{(d-2)d_{\perp}(d_\perp+2)}\,
                   \bar{f}\,,
 \label{zepatrani} \\
        &&\zeta_{\lambda}= -\frac{6(d_{\perp}-2)
               (d_{\perp}-1)}{(d-2)d_{\perp}(d_\perp+2)}\,\bar{f}\,,
 \label{zetlambdani} \\
        &&\zeta_D= -\frac{1}{4(d_{\perp}+2)}\,\bar{f}\,,
 \label{zetdptani} \\
        &&\zeta_{\tau}= -2+\frac{1}{d+d_{\parallel}-2}\,v\,-\,
       \frac{2(d_{\perp}-1)}{d_{\perp}}\\
        &&\mbox{}\times\left(\frac{1}{d+d_{\parallel}-2}
       -\frac{3(d_{\perp}-2)}{(d-2)(d_{\perp}+2)}\right)\,\bar{f}\,,
 \nonumber
 \label{zetrptani} \\      
        &&\zeta_{\widetilde{u}} = -\epsilon+
  \frac{3}{2}\,v-\frac{2(d_{\perp}-1)}{(d-2)d_{\perp}}\,f\\
&&\mbox{}-\frac{d_{\perp}-1}{d_{\perp}}
       \,\left(1-\frac{12(d_{\perp}-2)}{(d-2)(d_{\perp}+2)}\right)\,\bar{f}\,,
\nonumber 
\label{zetuptani}\\
&&2 \zeta_{\widetilde{g}}=d+d_\parallel-6-\frac{4}{(d-2)d_{\perp}}\,
\left(d_{\perp}-1-\frac{2}{d_{\perp}+2}\right)\,f\nonumber\\
&&\mbox{}+\frac{4}{d_{\perp}(d_{\perp}+2)}\,
\left(\frac{3(d_{\perp}-2)(d_{\perp}-1)}{d-2}
-\frac{d_{\perp}+16}{16}\right)\,\bar{f}\nonumber\\
&&\mbox{}+\frac{d}{4d_{\perp}(d_{\perp}+2)}\,\widetilde{f}\,,
\label{zetildegani}
\end{eqnarray}
with $\zeta_c=-\zeta_{\lambda}$ from Eq.~(\ref{Zlc}) and 
$2\zeta_g=d+d_\parallel-6-\zeta_{\scj}$ from (\ref{ZGalani}). These equations
reduce to the equilibrium ones in the limit $d_\parallel=0$, i.e., 
$d_{\perp}=d$.

In the anisotropic two-temperature model H, the scaling laws 
(\ref{opcorr})--(\ref{opsusc}) generalize to
\begin{eqnarray}
  C_\psi(\tau,{\bf q}_\parallel,{\bf q}_\perp,\omega) &=& 
  q_\perp^{-2-{\widetilde z}_\psi+\eta_\psi} \, {\hat C_\psi}\left( 
  \frac{\tau}{q_\perp^{1/\nu}},\frac{q_\parallel}{q_\perp^{1+\Delta}},
  \frac{\omega}{q_\perp^{z_\psi}}\right) , \nonumber \\ && \label{opcoan} \\
  C_{\scj}(\tau,{\bf q}_\parallel,{\bf q}_\perp,\omega) &=& 
  q_\perp^{-2-{\widetilde z}_{\scj}} \, {\hat C_{\scj}}\left( 
  \frac{\tau}{q_\perp^{1/\nu}},\frac{q_\parallel}{q_\perp^{1+\Delta}},
  \frac{\omega}{q_\perp^{z_{\scj}}}\right) , \label{jcoran} \\
  \chi(\tau,{\bf q}_\parallel,{\bf q}_\perp,\omega) &=& q_\perp^{-2+\eta} \, 
  {\hat \chi}\left( \frac{\tau}{q_\perp^{1/\nu}},
  \frac{q_\parallel}{q_\perp^{1+\Delta}},\frac{\omega}{q_\perp^{z_\psi}}\right)
  , \label{opsuan}
\end{eqnarray}
and the critical exponents are defined via
\begin{eqnarray}
\label{etaani}
\eta&=&-\zeta^*\,,\\
\eta_{\psi}&=&-\zeta^*_{\psi}\,,
\label{etapani}\\
\nu^{-1}&=&-\zeta_{\tau}^*\,,
\label{nuani}\\
\Delta&=&1-\frac{\zeta_c^*}{2}=1+\frac{\zeta_{\lambda}^*}{2}\,,
\label{Deltaani}\\
z_{\psi}&=&4+\zeta_{\lambda}^*\,,
\label{zpsiani}\\
z_{\scj}&=&2+\zeta_D^*\,,
\label{zjani}
\end{eqnarray}
where the exponent $\Delta$
originates from the intrinsic 
aniso\-tropy of the system \cite{bearev,bearoy,uwejai}.

From the zeta functions (\ref{zetlambdani})--(\ref{zetildegani}), one can 
compute the beta functions for the coupling constants $v$, $f$ and 
$\widetilde{f}$, which determine the fixed points, with $\bar{f}$ being given
by $\bar{f}=\sqrt{f\widetilde{f}}$. To one-loop order, these beta functions 
read
\begin{eqnarray}
\label{anibetav}
\beta_{v}&=&\left(\zeta_{\widetilde{u}}+\frac{d_{\parallel}}{2}\,
\zeta_{\lambda}\,\right)\,v\\
&=&\left[-\epsilon +\frac{3}{2}\,v
-\frac{2(d_\perp-1)}{(d-2)d_\perp}\,f\right.\nonumber\\
&&\mbox{}\left.-\frac{d_\perp-1}{d_\perp}\,\left(1+
\frac{3(d_\parallel-4)(d_\perp-2)}{(d-2)(d_\perp+2)}\right)\,\bar{f}
\right]v\,,\nonumber\\
\label{anibetaf}
\beta_{f}&=&\left[2(\zeta_{g}+1)-\zeta_{D}+
\left(\frac{d_{\parallel}}{2}-1\right)\,\zeta_{\lambda}\,\right]\,f=
-\epsilon\,f\nonumber\\
&&\mbox{}+\left[\left(\frac{1}{2(d_\perp+2)}
-\frac{3(d_\parallel-2)(d_\perp-2)(d_\perp-1)}{(d-2)d_\perp(d_\perp+2)}\right)
\,\bar{f}\right.\nonumber\\
&&\mbox{}\left.-\frac{d}{4d_\perp(d_\perp+2)}\,\widetilde{f}\right]\,f\,,
\\
\label{anibetaftilde}
\beta_{\widetilde{f}}&=&\left[2(\zeta_{\widetilde{g}}+1)-\zeta_{D}+
\left(\frac{d_{\parallel}}{2}-1\right)\right]\,\widetilde{f}\\
&=&\left[-\epsilon-\frac{4}{(d-2)d_\perp}\left(
d_\perp-1-\frac{2}{d_\perp+2}\right)f\right.\nonumber\\
&&\mbox{}-\frac{1}{d_\perp(d_\perp+2)}\left(4+
\frac{3(d_\parallel-6)(d_\perp-2)(d_\perp-1)}{d-2}\right)\,\bar{f}\nonumber\\
&&\mbox{}\left.+\frac{d}{4d_\perp(d_\perp+2)}\widetilde{f}\right]\,
\widetilde{f}\,,\nonumber
\end{eqnarray}
and it follows from $\bar{f}=\sqrt{f\widetilde{f}}$ that
\begin{equation}
\label{anibetabarf}
\beta_{\bar{f}}=\frac{1}{2}\,\left(\frac{\beta_{f}}{f}+
\frac{\beta_{\widetilde{f}}}{\widetilde{f}}\right)\,\bar{f}\,,
\end{equation}
which is used, together with (\ref{anibetaf}) and (\ref{anibetaftilde}),
to compute $\beta_{\bar{f}}$.

Since the beta functions $\beta_{f}$ and $\beta_{\widetilde{f}}$
do not depend on the static coupling $v$, the determination of the
fixed points reduces to the solution of the system of 
quadratic equations given by $\beta_{f}=\beta_{\widetilde{f}}=0$,
the equation for $\beta_{\bar{f}}$ being automatically 
satisfied. Introducing the following condensed notation
\begin{eqnarray}
\!\!\!\!a\!&=&\!\frac{2\,(d-2)\,d_{\perp}\,(d_{\perp}+2)}{(d-2)d_{\perp}
-6(d_{\parallel}-2)(d_{\perp}-2)(d_{\perp}-1)}\,,
\label{aqani}
\\
\!\!\!\!\alpha\!&=&\!\frac{4\,(d-2)\,d_{\perp}\,(d_{\perp}+2)}{(d\!-\!2)
(d_{\perp}\!-\!8)\!-\!12(d_{\parallel}\!
-\!4)(d_{\perp}\!-\!2)(d_{\perp}\!-\!1)}\,,
\label{alphaqani}
\\
\!\!\!\!c\!&=&\!\frac{(d-2)\,d}{2[(d-2)d_{\perp}
-6(d_{\parallel}-2)(d_{\perp}-2)(d_{\perp}-1)]}\,,
\label{cqani}
\\
\!\!\!\!\gamma\!&=&\!\frac{8\,[(d_{\perp}-1)(d_{\perp}+2)-2]}{(d\!-\!2)
(d_{\perp}\!-\!8)\!-\!12(d_{\parallel}\!
-\!4)(d_{\perp}\!-\!2)(d_{\perp}\!-\!1)}\,,
\label{gammaqani}
\end{eqnarray}
one can write the solutions of the system of quadratic
equations as
\begin{eqnarray}
f^\ast_{\pm}&=&\frac{\alpha-a-2\alpha\gamma c\pm
\sqrt{(a-\alpha)^2+4ac\alpha\gamma}}{2\,\gamma\,(\gamma c -1)}\,
\epsilon\,,
\label{ffani}\\
\widetilde{f}^\ast_{\pm}&=&
\frac{a-\alpha-2ac \gamma \pm
\sqrt{(a-\alpha)^2+4ac\alpha\gamma}}{2\,c\,(\gamma c -1)}\,
\epsilon\,.
\label{tilffani}
\end{eqnarray} 
The existence of fixed points depends on the existence of
at least one positive root for each of these two equations,
which gives a series of conditions on the coefficients
$a$, $c$, $\alpha$ and $\gamma$. However, it is immediate
to see from Eqs.~(\ref{ffani}) and (\ref{tilffani})
that, even when those solutions exist, they diverge if $\gamma c=1$.
Taking into account that $d=d_\perp+d_\parallel$, this
condition defines a sixth-order equation determining the relation
between $d_\perp$ and $d_\parallel$. With the minimal
subtraction prescription that $d=4-d_\parallel$, i.e., setting
$\epsilon=0$ in this equation, one obtains the numerical
solution $d_\parallel=0.838454$; already for $0\leq d_\parallel=1$, 
the RG flow takes the mode coupling to infinity. This result is completely
analogous to the result we have previously obtained in our study of the
two-temperature non-equilibrium model J \cite{uwejai}.

Formally, and following our study of model J, 
we may expand about the equilibrium model H, and thus obtain
critical exponents in the limit $d_\parallel \ll 1$.
To first order in $d_\parallel \epsilon$, we find
\begin{eqnarray}
        f^\ast&=&\frac{24}{19}\,\epsilon + \frac{1442}{6137}
                    \,d_\parallel\epsilon\,,
\label{fexpani}\\
        {\widetilde f}^*&=&\frac{24}{19}\,\epsilon + 
        \frac{11246}{6137}\,d_\parallel\,\epsilon\,,
\label{tfexpani}\\
        v^* &=& \frac{2}{3}\,\epsilon + \frac{143}{323}\,
d_\parallel\epsilon\,,
\label{vexpani}
\end{eqnarray}
leading to the critical exponents
\begin{eqnarray}
        &&\eta = -\frac{12}{19}\,d_\parallel\epsilon\; , \quad
        \eta_\psi = -\frac{21}{646}\,d_\parallel\epsilon\,,
 \label{mhexet} \\
        &&\nu^{-1} = 2 - \frac{1}{3}\,\epsilon-\frac{41}{646}\,d_\parallel
          \epsilon\, ,
 \label{mhexst} \\
        &&z = 4 - \frac{18}{19}\,\epsilon-\frac{2820}{6137}\,d_\parallel
          \epsilon\, ,
 \label{mhexdy} \\
         &&z_{\scj} = 2- \frac{1}{19}\,\epsilon-\frac{372}{6137}\,d_{\parallel}
          \epsilon\, ,
\label{mhexdyd}\\
        &&\Delta = 1 - \frac{9}{19}\,\epsilon-\frac{1410}{6137}\,d_\parallel
          \epsilon\, .
 \label{mhexcr}
\end{eqnarray}

Notice that this procedure amounts to an expansion with respect to 
{\em two} dimensional parameters, namely $\epsilon = 4 - d - d_\parallel$,
and $d_\parallel \epsilon$.
Moreover, the divergence of the non-expanded fixed point ${\widetilde f}^*$ at
$d_\parallel \approx 0.838454$ indicates that an extrapolation of the formal
results (\ref{mhexet}) to (\ref{mhexcr}) to any physical dimension 
$d_\parallel \geq 1$ is unlikely to work. On the other hand, we cannot exclude 
that, also for model H, this divergence merely represents an one-loop
artifact, and is cured if one calculates the RG beta functions to higher loop 
orders. Yet another possibility might well be that the divergence of 
$f^*$, ${\widetilde f}^*$
and $v^*$ indicates the absence of a {\em simple} non-equilibrium stationary 
state of the two-temperature model H in the vicinity of its critical point.
For example, in a {\em uniformly} rather than randomly driven non-equilibrium
version of model J, a similar divergence has been found recently \cite{dasrao}.
In that case, computer simulations have revealed that the system enters a 
regime of spatio-temporal chaos at long times; perhaps the absence of a finite 
RG fixed point in the randomly driven two-temperature models J and H might 
indicate similar behavior.
A somewhat less drastic implication may be that merely perturbation theory
breaks down, and non-perturbative approaches could possibly characterize the 
scaling behavior at the transition of the two-temperature model H successfully.

\section{Summary and final remarks}
 \label{concl}

We have studied two non-equilibrium generalizations of the dynamical
model H with both conserved scalar order parameter and dynamically coupled
conserved transverse currents that describes second-order liquid-gas or 
binary-fluid phase transitions. Specifically, we were interested in the 
effect of detailed balance violations on the asymptotic critical behavior.
We have investigated both (a) isotropic violations of the equilibrium
conditions, which can be formulated in terms of different effective
noise temperatures for the order parameter and conserved currents,
respectively, and (b) spatially anisotropic detailed balance violations, 
i.e., dynamical noise which is governed by different strenghts in
longitudinal and transverse momentum space sectors. 

In principle, there are several possible scenarios: (1) In the vicinity
of the critical point, detailed balance may effectively become restored
as a consequence of the diverging correlation length that essentially
averages over the different local noise temperatures; (2) a novel, stable
renormalization group fixed point may emerge that describes a new universality
class with genuine non-equilibri\-um scaling behavior; (3) there might be
no stable RG fixed point at all, indicating perhaps complex spatio-temporal
chaotic behavior rather than a simple stationary non-equi\-librium state.

We find that situation (1) applies to the isotropic non-equilibrium model H,
while scenario (3) appears to describe the effective two-temperature model
that emerges upon allowing for spatially anisotropic noise correlations.
Remarkably, case (2) is never realized in any dynamical model with reversible
mode-couplings. In fact, a surprisingly simple overall picture emerges
(we have already presented a brief overview in Ref.~\cite{uwevam}). Namely,
quite generally, the equilibrium dynamical models as listed in 
Ref.~\cite{hohhal} with {\em non-conserved} order parameter turn out to be
quite robust against detailed-balance violations. The purely relaxational
models A and C do not even have a genuine non-equilibrium fixed point at all.
For the SSS model, generalizing models E (for planar ferromagnets, $n = 2$)
and model G (for isotropic antiferromagnets, $n = 3$) to arbitrary order
parameter space dimension $n$, two non-equilibrium fixed points do exist,
corresponding to ratios $T = 0$ and $T = \infty$ for the noise temperatures
of the order parameter and dynamically coupled conserved fields, but neither
of these is stable. Thus, near the critical point, detailed balance becomes
eventually restored, and the asymptotic critical exponents are those of the
equilibrium model \cite{uwezol}. Of course, such systems might remain in 
the crossover region for quite a while, masking the asymptotic regime. 
Essentially, this scenario (1) also applies when the conserved noise for 
the coupled non-critical fields is rendered anisotropic as well. Additional 
fixed points emerge, but the isotropic equilibrium one remains stable
\cite{uwejai}.

When detailed balance is violated {\em isotropically} in the models B, D
(purely relaxational), J (isotropic ferromagnets, $n=3$) and H with
{\em conserved} order parameter, basically the same statements apply, and
scenario (1) is realized again \cite{uwezol,uwevam}. However, once one 
allows for spatially {\em anisotropic} or dynamical noise, separating a soft 
transverse and stiff longitudinal momentum space sector, which enforces 
anisotropic scaling, the behavior changes dramatically. In the relaxational 
two-temperature models B and D, the ensuing asymptotic theory however turns 
out to be equivalent to an equilibrium model with {\em long-range} 
correlations of the uniaxial dipolar or elastic type \cite{bearoy,uwevam}. 
This corresponds to case (2) above. In stark contrast, in the anisotropic
non-equilibrium versions of models J \cite{uwejai} and H which are 
characterized by relevant reversible mode-couplings to additional 
conserved variables, no stable renormalization group fixed point can be 
found (at least to one-loop order), which represents scenario (3).
We do at this point not really know what the absence of an RG fixed point
means physically in this situation; perhaps, as in the uniformly driven
non-equilibrium model J \cite{dasrao}, the long-time behavior is governed
by spatio-temporal chaos. It would certainly be worthwhile to explore this
issue further, e.g., through computer simulations.

\begin{acknowledgement}
J.E.S. acknowledges support from the Deutsche Forschungsgemeinschaft
(SFB 413/TP C6) and the European Commission through a Marie Curie 
TMR fellowship (ERB-FMBI-CT 97-2816). U.C.T. is grateful for support  
by the National Science Foundation through the Division of Materials 
Science (grant no. DMR-0075725), as well as through the Jeffress 
Memorial Trust (grant no. J-594).
We benefited from helpful discussions with E. Frey, H.K. Janssen,
Z. R\'acz, B. Schmittmann, F. Schwabl, and R.K.P. Zia.
\end{acknowledgement}

\appendix
\section{Galilean invariance in the dynamic model H}
\label{appA}

In this appendix we derive the basic Ward identity which was used in the 
analysis of the non-equilibrium isotropic and two-temperature model H, 
namely Eq.~(\ref{galZ}), which follows from the Galilean invariance of 
the Langevin equations which describe the model \cite{modelH}. In order 
to prove such an identity, we write the source-free equations 
(\ref{hoplan}) and (\ref{samlan}) in a slightly different form:
\begin{eqnarray}
        {\partial \psi_0\over \partial t}+g_0\,\vecj_0\cdot\nabla\psi_0&=&
        \lambda_0\nabla^2(r_0-\nabla^2)\psi_0\nonumber\\
        &&\mbox{}+{\lambda_0 u_0\over 6}\nabla^2\psi^3_0+\eta\,,
        \label{hoplanA}
\end{eqnarray}
and
\begin{eqnarray}
        &&{\partial \mbox{\bf \j}_0\over \partial t} 
        +g_0\,\vecj_0\cdot\nabla\vecj_0 = \nonumber \\
        &&\quad = {\cal T}\left[D_0\nabla^2\vecj_0
            +g_0\nabla\psi_0(r_0-\nabla^2)\psi_0+\vec{\zeta}\right]\, , 
\label{samlanA}
\end{eqnarray}
where we have added the convective term $g_0\,\vecj_0\cdot\nabla\vecj_0$
to the left-hand side of (\ref{samlanA}). This term is normally not 
included explicitly in the analysis, as it generates diagrammatic 
contributions proportional to $g_0^2/D_0^2$. Since the scaling dimension 
of this effective coupling is $\mu^{2-d}$ ($\mu^{2-d-d_\parallel}$ for 
the two temperature model H), as can be seen from the dimensional 
analysis of sections \ref{verresp} and \ref{mdjren}, this coupling is 
irrelevant in the renormalization group sense and therefore normally 
neglected.

Under a Galilean transformation, to a reference frame moving with 
respect to the laboratory frame with velocity $\vec{v}$, the coordinates 
and fields change according to
\begin{eqnarray}
\vec{r}'&=&\vec{r}-g_0\,\vec{v}\,t\,,
\label{Galr}\\
t'&=&t\,,
\label{Galt}\\
\psi_0'(\vec{r}',t')&=&\psi_0'(\vec{r}-g_0\,\vec{v}\,t,t)=\psi_0(\vec{r},t)\,,
\label{Galpsi}\\
\vecj_0'(\vec{r}',t')&=&\vecj_0'(\vec{r}-g_0\,\vec{v}\,t,t)=\vecj_0(\vec{r},t)
-\vec{v}\,,
\label{Galj}
\end{eqnarray}
where the prime ($'$) refers to parameters and dynamic variables
measured in the moving frame. 

Using these transformation laws, it is easy to show that the time and
space derivatives of $\psi_0$ and $\vecj_0$ are transformed according to
\begin{eqnarray}
\nabla_{\vec{r}'}\mid_{t'}&=&\nabla_{\vec{r}}\mid_{t}\,,
\label{Galnabla}\\
\frac{\partial}{\partial t'}\mid_{\vec{r}'}&=&
\frac{\partial}{\partial t}\mid_{\vec{r}}+
g_0\,\vec{v}\cdot\nabla_{\vec{r}}\mid_{t}\,,
\label{Galpartdt}
\end{eqnarray}
where $\mid_{\vec{r}}$ etc. simply indicates which variable is being
held constant when the derivative is taken. With these relations, it
is easy to show that the material derivative $d / dt$ which appears 
on the left-hand side of (\ref{hoplanA}) and (\ref{samlanA}), i.e.
\begin{eqnarray}
\frac{d}{dt}&=&\frac{\partial}{\partial t}\mid_{\vec{r}}+
g_0\,\vecj_0\cdot\nabla_{\vec{r}}\mid_{t}\nonumber\\
&=&\frac{\partial}{\partial t'}\mid_{\vec{r}'}
  +g_0\,\vecj_0'\cdot\nabla_{\vec{r'}}\mid_{t'}\,,
\label{Galmat}
\end{eqnarray}
is invariant under a Galilean transformation, i.e. it preserves its 
form on going from one reference frame to another, as indicated in 
Eq.~(\ref{Galmat}). The right-hand side of Eqs.~(\ref{hoplanA}) and 
(\ref{samlanA}) can also be seen from Eqs.~(\ref{Galpsi}) to 
(\ref{Galnabla}) to be trivially invariant, given the fact that 
$\vec{v}$ is a constant vector. It is thus shown that the Langevin 
equations describing model H are invariant under a Galilean 
transformation (the distribution of the noise being the same in both 
reference frames).

This invariance must be preserved under renormalization, i.e., when we 
substitute $\psi_0$, $\scj_0$, $g_0$, $\ldots$, by their renormalized 
counterparts $\psi$, $\scj$, $g$, etc. For this to happen, the 
renormalization factors $Z_g$ and $Z_{\scj}$ have to compensate each 
other in Eq.~(\ref{Galmat}), i.e., one must have $Z_gZ_{\scj}\equiv 1$, 
which is Eq.~(\ref{galZ}). Notice that the same reasoning also applies 
to the two-temperature model H, once $g_0$ is substituted by 
$g_0=\sqrt{{\widetilde D}_0^\perp / D_0^\perp} \, g_0^\perp$.


\begin{thebibliography}{}
\bibitem{nonoether} Note that since we are considering stochastic dynamics on
		a coarse-grained description level, these conservation laws do 
	        not necessarily just correspond to a symmetry of the 
 		effective Hamiltonian governing the static critical properties.
\bibitem{orderparani} Note that in the cases usually discussed in static
  		 critical phenomena, anisotropy is introduced in the 
		 order parameter space, e.g. by adding an easy axis energy 
		 term to the Hamiltonian. It is the very absence of 
	         detailed-balance constraints that gives us this additional
	         freedom of introducing spatial anisotropy in the noise
	         correlations here.
\bibitem{eqth}   In the equilibrium theory, this is a very natural choice,
		 as the transverse currents are non-critical modes. Thus 
 		 their static response is only weakly temperature-dependent
		 and essentially constant near the critical point.
\bibitem{conT0}  Note that in Ref.~\cite{uwezol} we have defined, for 
		 convenience, $T_0$ as the inverse of Eq.~(\ref{eqT0}) here. 
	         The subsequent RG analysis is of course independent of 
		 this choice.
\bibitem{IRfint} In some of the integrals to be evaluated one obtains
                 contributions of the form $(\tau_0+D_0/\lambda_0)^{(d-4)/2}$.
                 In the IR limit $\tau_0\rightarrow 0$, these contributions
                 are finite below four dimensions and may be neglected,
                 following the minimal subtraction prescription. 
                 A more expedite way to deal with these terms simply consists 
		 in taking the asymptotic limit $\lambda_0/D_0 \rightarrow 0$, 
		 keeping $T_0$ finite.
\bibitem{factorPout} The vertex functions and susceptibilities of the
                 variables $\widetilde{\vecj}$ and $\vecj$ are, of course, 
	         tensor functions. But given that these functions are purely 
	         transverse, one can write them as, e.g., 
                 $\Gamma_{0\,\widetilde{\scj}\,\scj}^{\alpha\beta}
                 (\vec{q},\omega)={\cal T}^{\alpha\beta}(\vec{q})\,
                 \Gamma_{0\,\widetilde{\scj}\,\scj} (\vec{q},\omega)$,
	         where ${\cal T}^{\alpha\beta}(\vec{q})$ denotes the 
	         transverse projector defined above. We have therefore only 
	         explicitly given the scalar part 
                 $\Gamma_{0\,\widetilde{\scj}\,\scj}(\vec{q},\omega)$ 
		 in the main text.
\bibitem{care1} Note that although $Z_\psi=Z$ to all orders of 
	        perturbation theory in equilibrium, the equality 
	        $Z_\psi=Z=1$ holds only to one-loop order, i.e., the
                susceptibility (and consequently the exponent $\eta$)
                acquires higher order corrections.
\bibitem{nocritslow} Since the conserved current is a non-critical
                quantity, it does not show any critical slowing-down.
\end{thebibliography}
\end{document}